\documentclass[5p]{elsarticle}

\usepackage{graphicx}
\usepackage{multirow}
\usepackage{lipsum} 
\usepackage[breaklinks=true]{hyperref} 
\usepackage[T1]{fontenc}
\usepackage[utf8]{inputenc}
\usepackage{textcomp} 
\usepackage{lmodern}
\graphicspath{{./figures/}}

\usepackage{amsmath}

\biboptions{sort&compress}

\usepackage[switch]{lineno}

\usepackage{xcolor}
\hypersetup{
    colorlinks,
    linkcolor={blue!80!black},
    citecolor={blue!80!black},
    urlcolor={blue!80!black}
}
\definecolor{hlcolor}{RGB}{209, 21, 7}

\newcommand{\loz}{L1$\mathrm{_0}$}
\newcommand{\lozfeni}{L1$\mathrm{_0}$~FeNi} 
\newcommand{\lozfept}{L1$\mathrm{_0}$~FePt} 

\newcommand{\etal}{\textit{et al.}}
\newcommand{\MJ}{MJ\,m$^{-3}$}

\newcommand{\st}{$^\mathrm{o}$}

\newcommand{\muB}{$\mu_\mathrm{B}$}

\usepackage{etoolbox}

\makeatletter 
\def\@author#1{\g@addto@macro\elsauthors{\normalsize%
    \def\baselinestretch{1}%
    \upshape\authorsep#1\unskip\textsuperscript{%
      \ifx\@fnmark\@empty\else\unskip\sep\@fnmark\let\sep=,\fi
      \ifx\@corref\@empty\else\unskip\sep\@corref\let\sep=,\fi
      }%
    \def\authorsep{\unskip,\space}%
    \global\let\@fnmark\@empty
    \global\let\@corref\@empty  
    \global\let\sep\@empty}%
    \@eadauthor={#1}
}
\makeatother

\makeatletter
\g@addto@macro\@floatboxreset\centering
\makeatother

\begin{document}
\begin{sloppypar}

\title{
Magnetic anisotropy of L1$_0$ FeNi (001), (010), and (111) ultrathin films:\\ 
A first-principles study
}

\author{Joanna Marciniak}
\author{Miros\l{}aw Werwi\'nski\corref{cor1}}
\ead{werwinski@ifmpan.poznan.pl}
\cortext[cor1]{Corresponding author}
\address{Institute of Molecular Physics, Polish Academy of Sciences,  M. Smoluchowskiego 17, 60-179 Pozna\'n, Poland}

\begin{abstract}

%
In previous experiments, \lozfeni{} thin films with different surfaces, including (001), (110), and (111), were produced and studied.
Each surface defines a different alignment of the crystallographic tetragonal axis with respect to the film's plane, resulting in different magnetic anisotropies.
In this study, we use density functional theory calculations to examine three series of \lozfeni{} films with surfaces (001), (010), and (111), and with thicknesses ranging from 0.5 to 3~nm (from 4 to 16 atomic monolayers).
Our results show that films (001) have perpendicular magnetic anisotropy, while (010) favor in-plane magnetization, with a clear preference for the tetragonal axis [001].
We proposed calling this type of in-plane anisotropy \textit{fixed in-plane}. 
A film with a surface (111) and a thickness of four atomic monolayers has a magnetization easy axis almost perpendicular to the plane of the film. 
As the thickness of the (111) film increases, the direction of magnetization rotates towards a tetragonal axis [001], positioned at an angle of about 45$^{\circ}$ to the plane of the film.
Furthermore, the most significant changes in spin and orbital magnetic moments occur at a depth of about three near-surface atomic monolayers. 
The presented results could be useful for experimental efforts to synthesize ultrathin \lozfeni{} films with different surfaces. 
Ultrathin \lozfeni{} films with varying magnetic anisotropies may find applications in spintronic devices.

\end{abstract}

\maketitle

\section{Introduction}

\begin{table*}[!t]
\caption{
Structural data of \lozfeni{} (001), (010), and (111) thin films consisting of 16 atomic monolayers.
The $c$ parameters of the computational unit cells include a vacuum of at least 30~\AA{}.
The thicknesses (without vacuum) of the considered (001), (010), and (111) films are 27, 27, and 31~\AA{}, respectively.
Unit cell parameters ($a$, $b$, $c$) are given in \AA{}.
}
\vspace{1mm}
\label{tab-wyckoff-positions}
\begin{tabular}{c | c c c | c c c | c c c}
        \hline
        \hline
        \multirow{4}{*}{} & \multicolumn{3}{c}{(001)} & \multicolumn{3}{|c}{(010) } & \multicolumn{3}{|c}{(111)}\\
         & \multicolumn{3}{c}{s.g. $P$4$mm$}     &\multicolumn{3}{|c}{s.g. $Pmma$} &\multicolumn{3}{|c}{s.g. $P$12/$m$1}\\
        & $a$ & $b$ & $c$      & $a$ & $b$ & $c$   & $a$ & $b$ & $c$\\
   Atom  &  2.52    &  2.52    & 56.85      &  3.56    &  3.58    & 56.70&  2.52    &  4.38    & 60.89\\
        \hline
   Fe    &  0.00    &  0.00    & -0.484   &  0.75    &  0.00    &  0.484 &  0.00 &   0.335 &  -0.483\\
   Ni    &  0.50    &  0.50    &  0.484   &  0.25    &  0.50    &  0.484 &  0.50 &  -0.165 &  -0.483\\
   Fe    &  0.00    &  0.00    &  0.453   &  0.25    &  0.00    &  0.453 &  0.00 &   0.004 &  -0.449\\
   Ni    &  0.50    &  0.50    & -0.453   &  0.75    &  0.50    &  0.453 &  0.50 &  -0.496 &  -0.449\\
   Fe    &  0.00    &  0.00    & -0.422   &  0.75    &  0.00    &  0.422 &  0.00 &  -0.327 &  -0.416\\
   Ni    &  0.50    &  0.50    &  0.422   &  0.25    &  0.50    &  0.422 &  0.50 &   0.173 &  -0.416\\
   Fe    &  0.00    &  0.00    &  0.390   &  0.25    &  0.00    &  0.391 &  0.00 &   0.343 &  -0.382\\
   Ni    &  0.50    &  0.50    & -0.390   &  0.75    &  0.50    &  0.391 &  0.50 &  -0.157 &  -0.382\\
   Fe    &  0.00    &  0.00    & -0.359   &  0.75    &  0.00    &  0.360 &  0.00 &   0.012 &  -0.348\\
   Ni    &  0.50    &  0.50    &  0.359   &  0.25    &  0.50    &  0.360 &  0.50 &  -0.488 &  -0.348\\
   Fe    &  0.00    &  0.00    &  0.328   &  0.25    &  0.00    &  0.328 &  0.00 &  -0.321 &  -0.315\\
   Ni    &  0.50    &  0.50    & -0.328   &  0.75    &  0.50    &  0.328 &  0.50 &   0.180 &  -0.315\\
   Fe    &  0.00    &  0.00    & -0.296   &  0.75    &  0.00    &  0.297 &  0.00 &   0.353 &  -0.281\\
   Ni    &  0.50    &  0.50    &  0.296   &  0.25    &  0.50    &  0.297 &  0.50 &  -0.147 &  -0.281\\
   Fe    &  0.00    &  0.00    &  0.265   &  0.25    &  0.00    &  0.266 &  0.00 &   0.010 &  -0.248\\
   Ni    &  0.50    &  0.50    & -0.266   &  0.75    &  0.50    &  0.266 &  0.50 &  -0.489 &  -0.248\\ 
        \hline
        \hline
    \end{tabular}
\end{table*}

\begin{figure}[!t]
\hspace{0mm} \includegraphics[width=0.91\columnwidth]{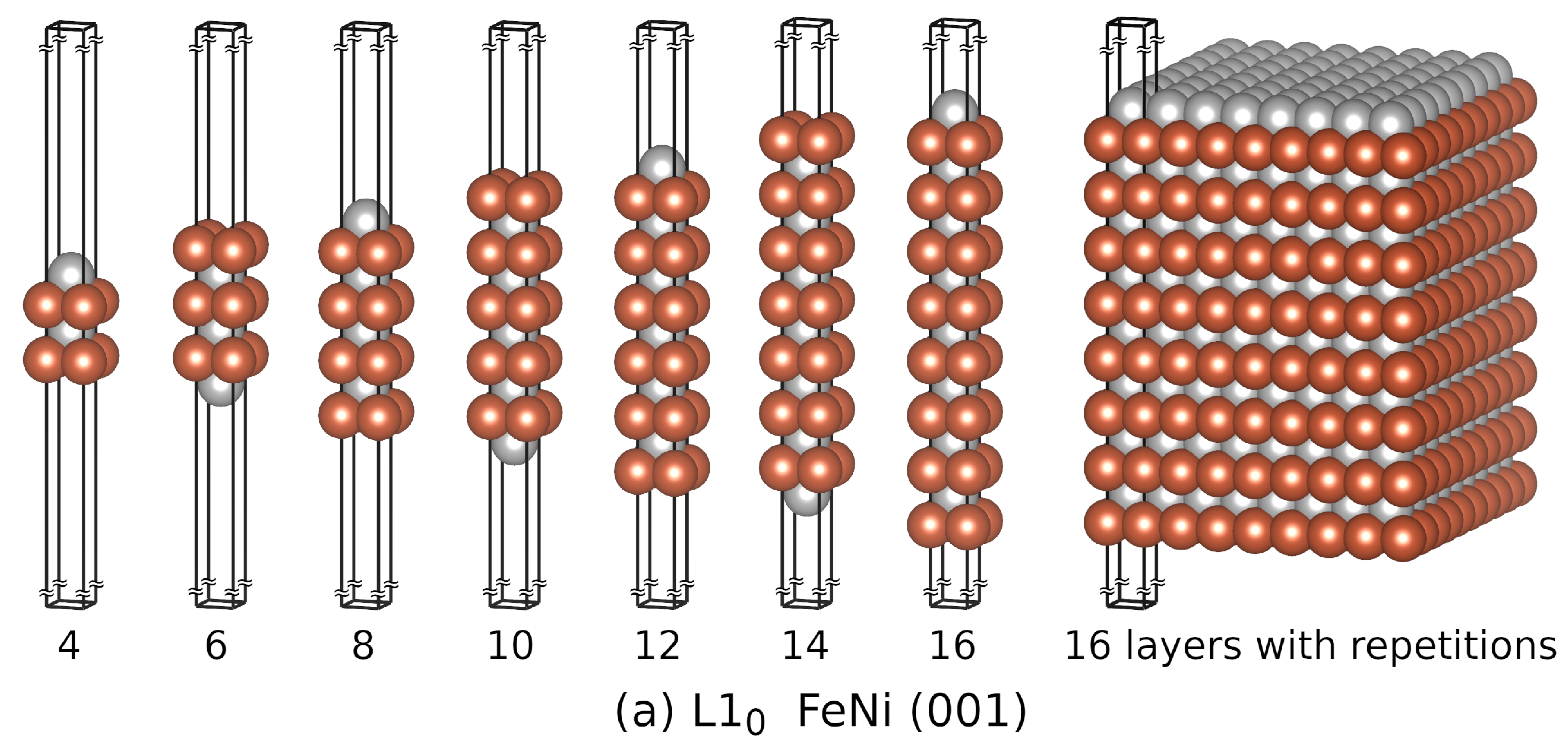}\hfill
\vspace{3mm}

\includegraphics[width=\columnwidth]{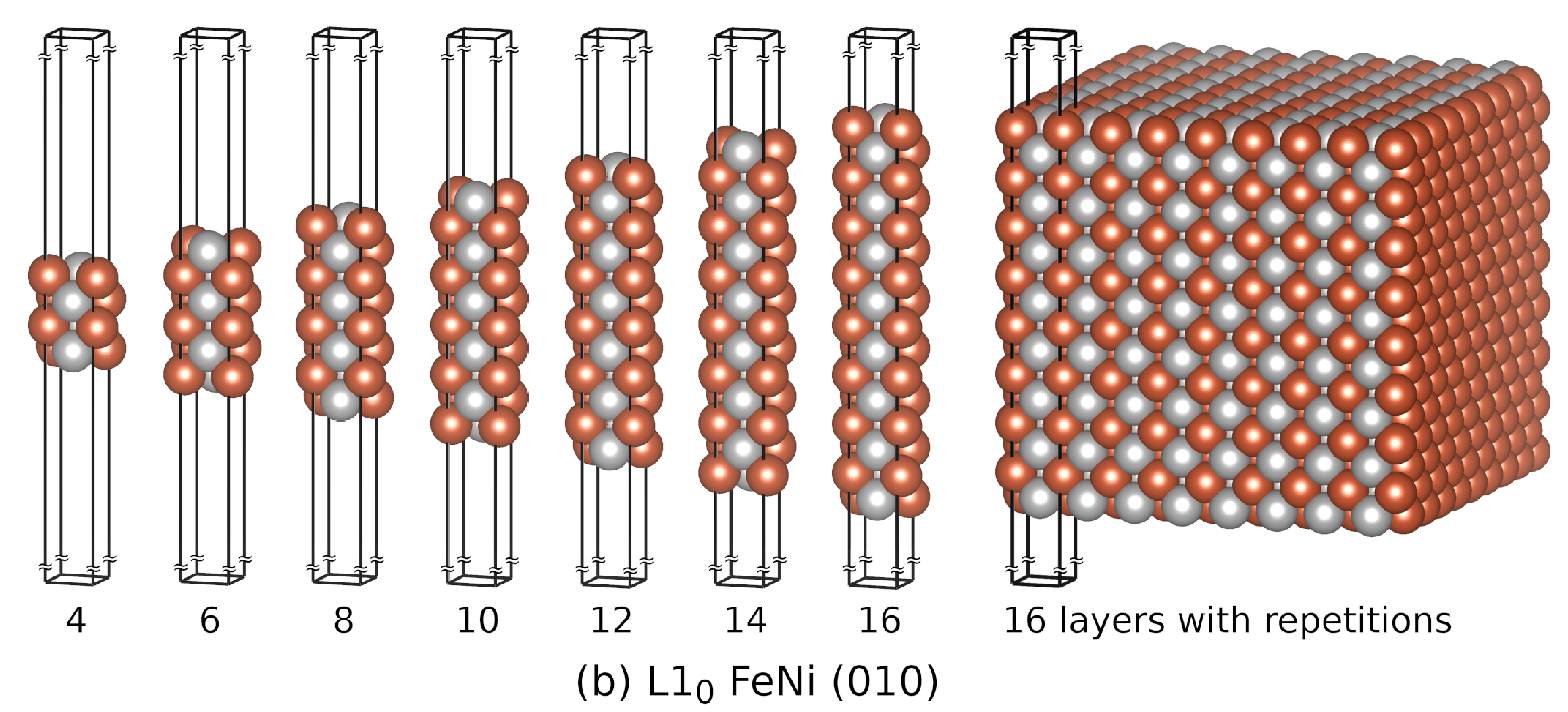}
\vspace{0mm}

\includegraphics[width=\columnwidth]{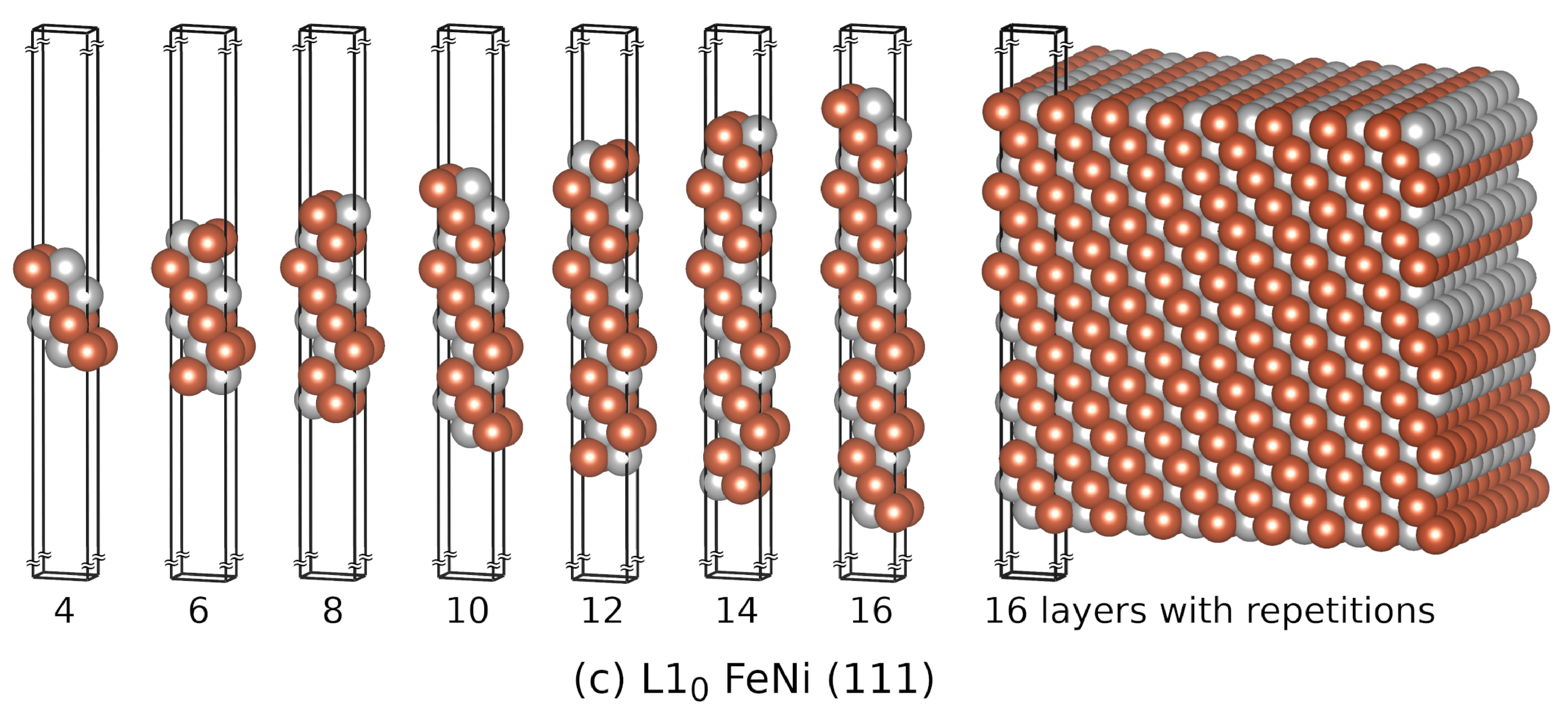}

\caption{\label{fig-structures}
The series of unit cells of \lozfeni{} thin films with (001), (010), and (111) surfaces and thickness from 4 to 16 atomic monolayers
[from 0.5 to 2.7~nm for (001) and (010) films and from 0.6 to 3.1~nm for (111) films]. 
On the right, 16-monolayer unit cells were repeated several times in two directions in the plane of the film to visualize the layered character of the models.
The vacuum height (at least 30~\AA{}) has been reduced in the figures to save space.
}
\end{figure}

%
\begin{figure}[!t]

\includegraphics[clip,width=1.0\columnwidth]{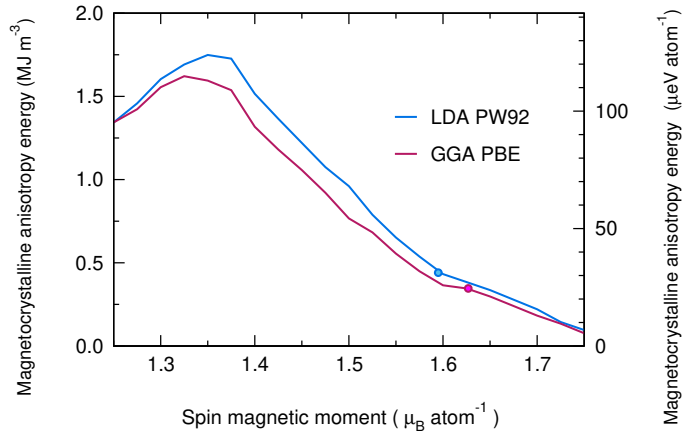}
\caption{\label{fig-bulk_MAE_vs_fsm}
Magnetocrystalline anisotropy energy of the bulk \lozfeni{} as a function of the spin magnetic moment.
Calculations were performed using the FPLO18 code with the 
LDA PW92 and GGA PBE exchange-correlation potentials.
Circles represent equilibrium results.
}
\end{figure}

%
\begin{figure*}[!t]
\hfill
\includegraphics[trim = 0 0 0 0, clip,width = 0.295\textwidth]{feni_bulk_scalar_pbe_dos.eps}
\includegraphics[trim = 143 36 85 90, clip,width = 0.49\textwidth]{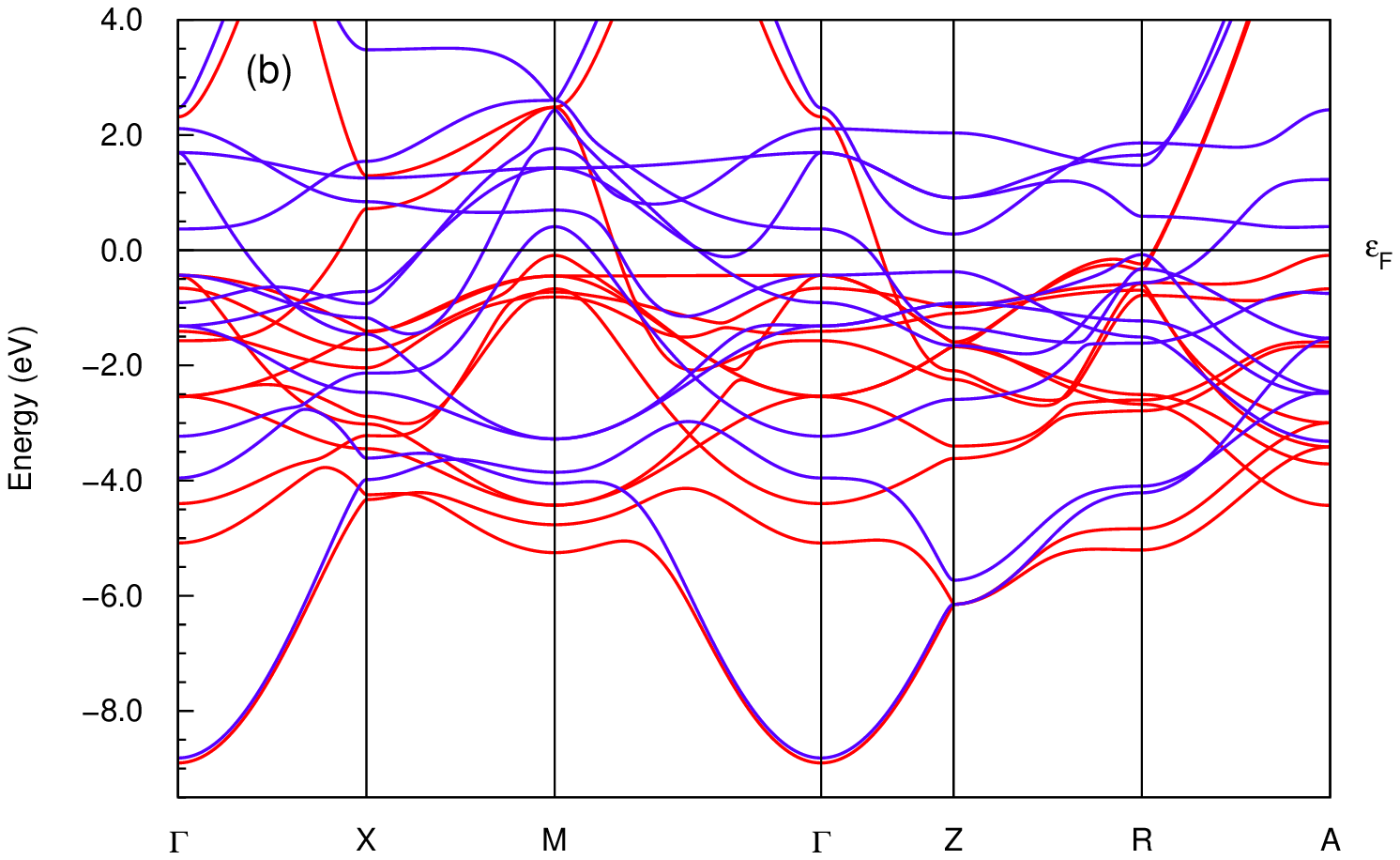}
\hfill
\includegraphics[trim = 0 -400 0 0,clip,width = 0.16\textwidth]{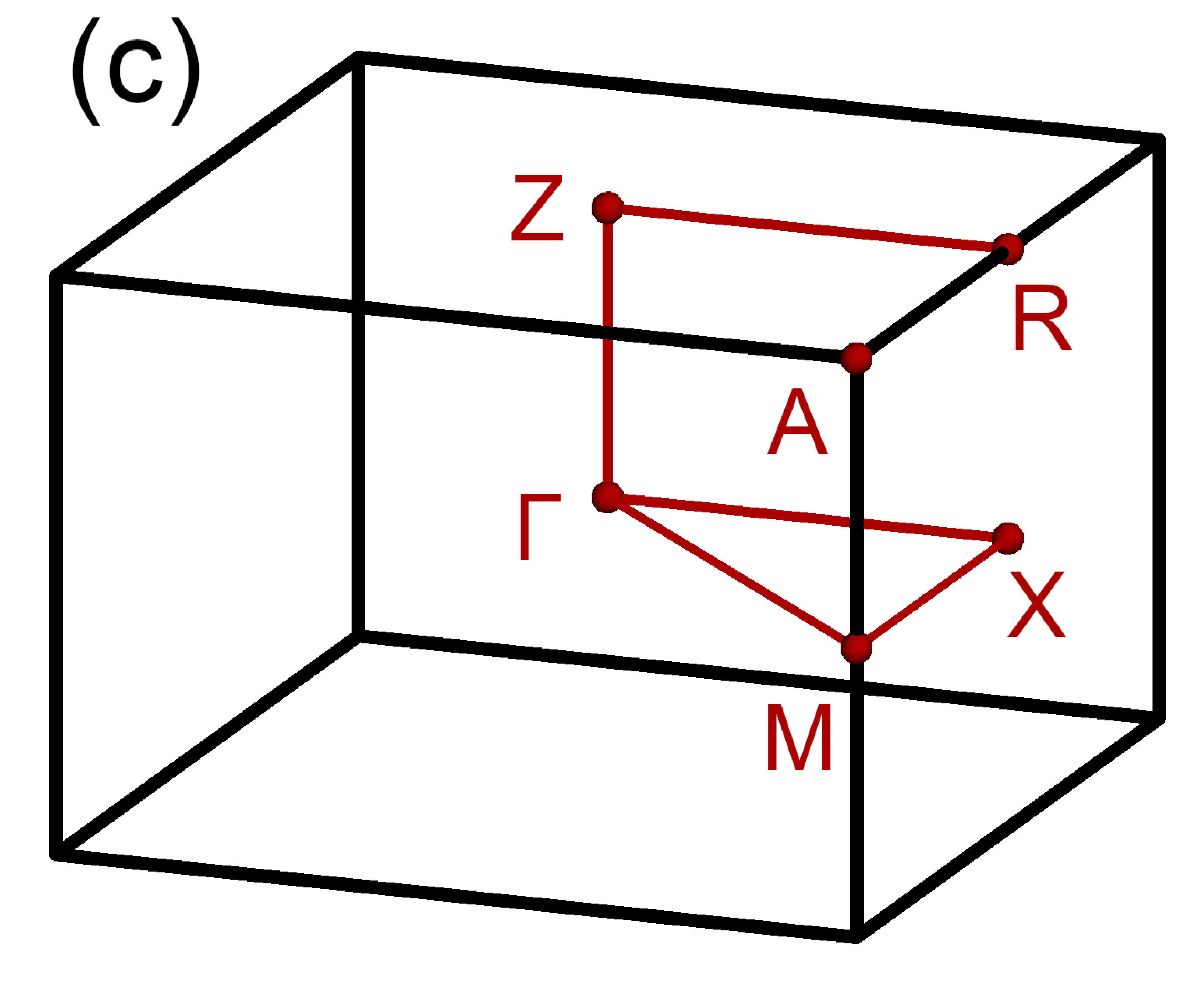}
\hfill
\caption{\label{fig:bulk_dos_bands_bzone}
Density of states (a) and band structure (b) of \lozfeni{} bulk.
Red and blue bands denote contributions from majority and minority spin channels, respectively.
Calculations were performed with the FPLO18 code using the PBE functional and a scalar-relativistic approach. 
Brillouin zone with high symmetry points is also shown (c).
}
\end{figure*}

%
\begin{figure}[t]
\includegraphics[clip,width=0.97\columnwidth]{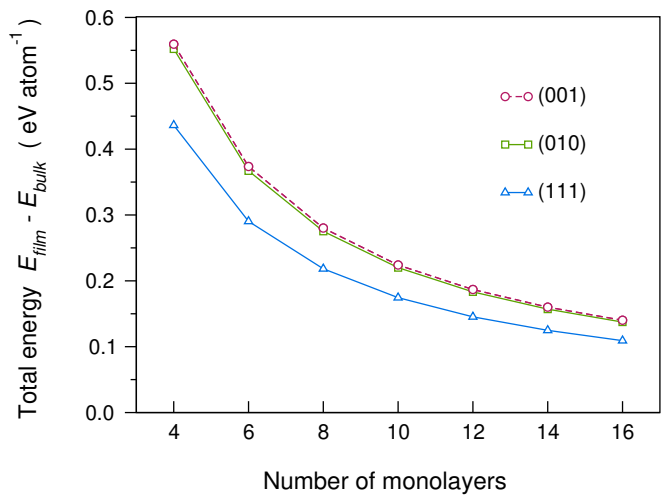}\hfill{}
\caption{\label{fig-energy_vs_nml}
Differences between film energy and bulk energy, calculated for \lozfeni{} films with (001), (010), and (111) surfaces.
Calculations were performed using the FPLO18 code with the PBE exchange-correlation functional.
}
\end{figure}

%
\begin{figure}[!t]
\includegraphics[trim = 0 5 0 12,clip,width=0.835\columnwidth]{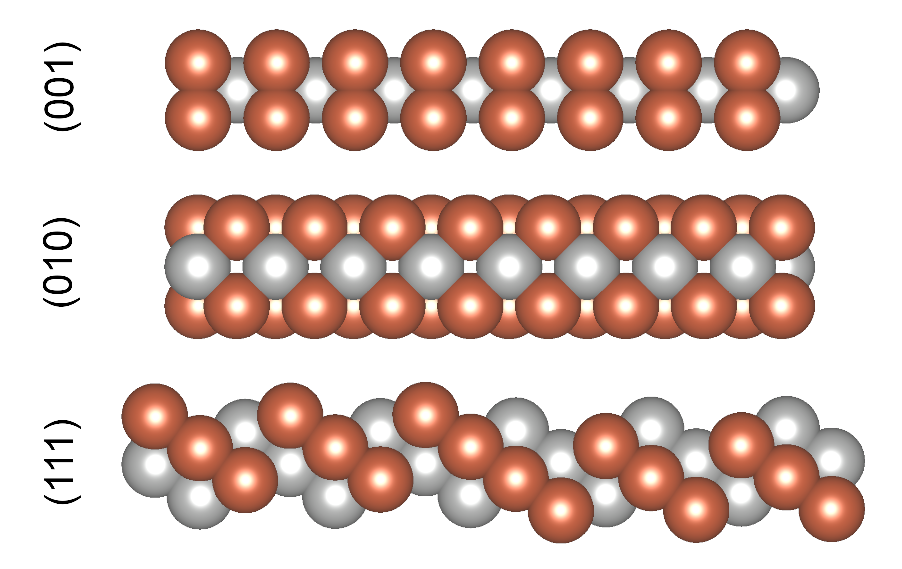}
\includegraphics[trim = 0 26 0 0,clip,width=0.825\columnwidth]{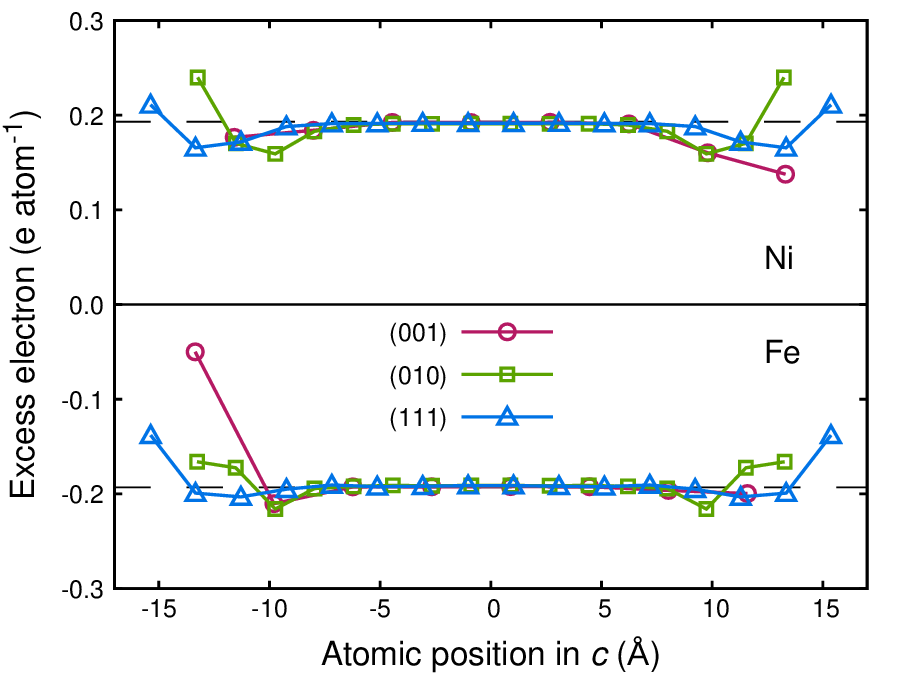}
\includegraphics[trim = 0 26 0 0,clip,width=0.835\columnwidth]{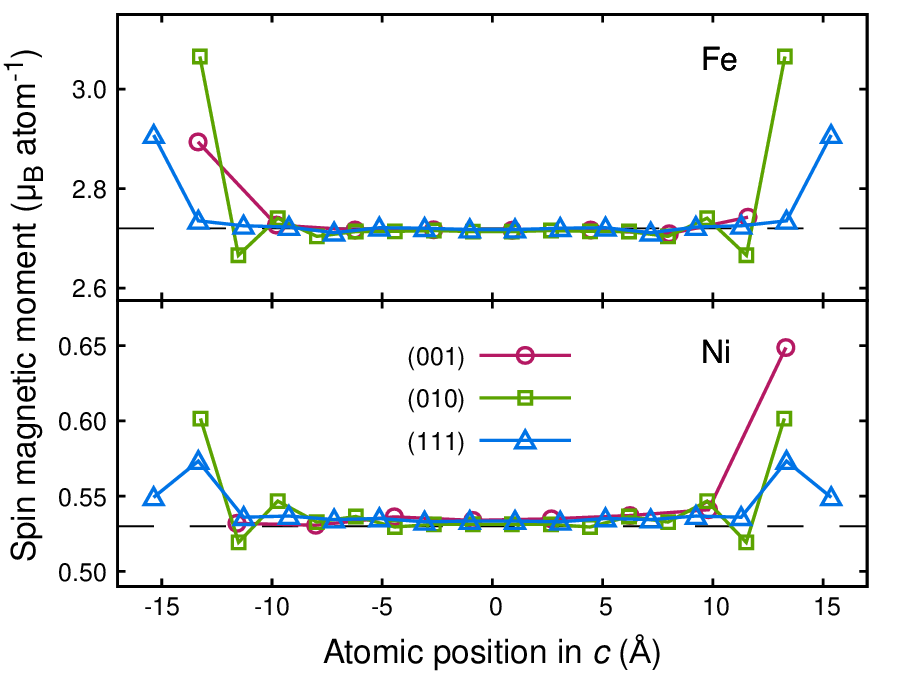}
%
\includegraphics[trim = 0 0 0 0,clip,width=0.835\columnwidth]{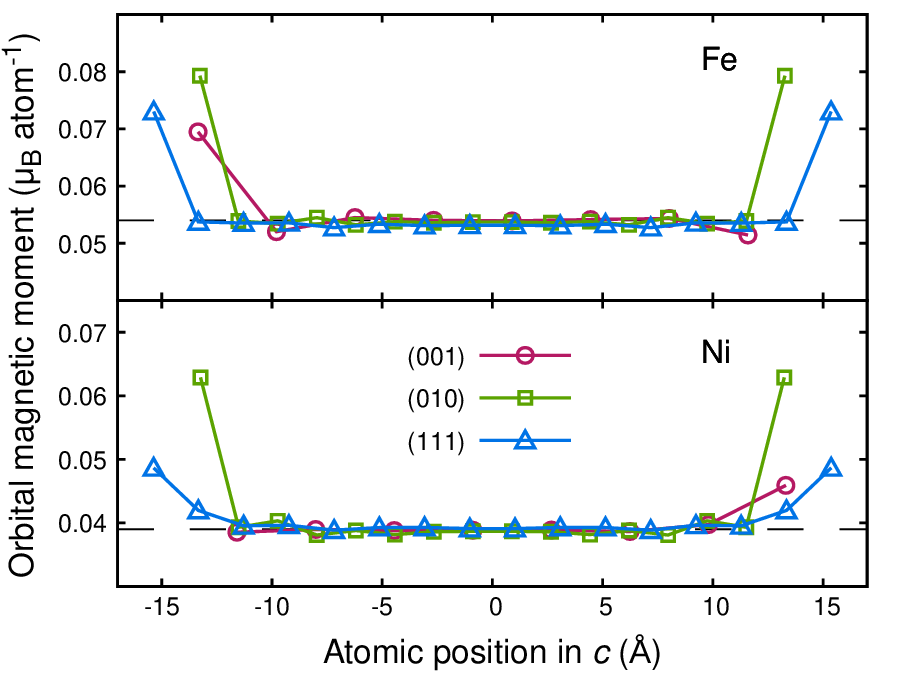}
\caption{\label{fig-m_vs_position}
Excess electrons and spin and orbital magnetic moments in 16-monolayer films of \lozfeni{} with surfaces (001), (010), and (111). 
Dashed lines indicate bulk values: spin magnetic moments on Fe (Ni) equal to 2.72~$\mu_\mathrm{B}$ (0.53~$\mu_\mathrm{B}$), orbital magnetic moments on Fe (Ni) equal to 0.054~$\mu_\mathrm{B}$ (0.039~$\mu_\mathrm{B}$), and excess electron of $\pm$0.193~$e$.
Calculations were performed using the FPLO18 code with the PBE exchange-correlation potential.
Above the charts are the side views of the film unit cells.
}
\end{figure}

%
\begin{figure}[t]

\includegraphics[clip,width=0.95\columnwidth]{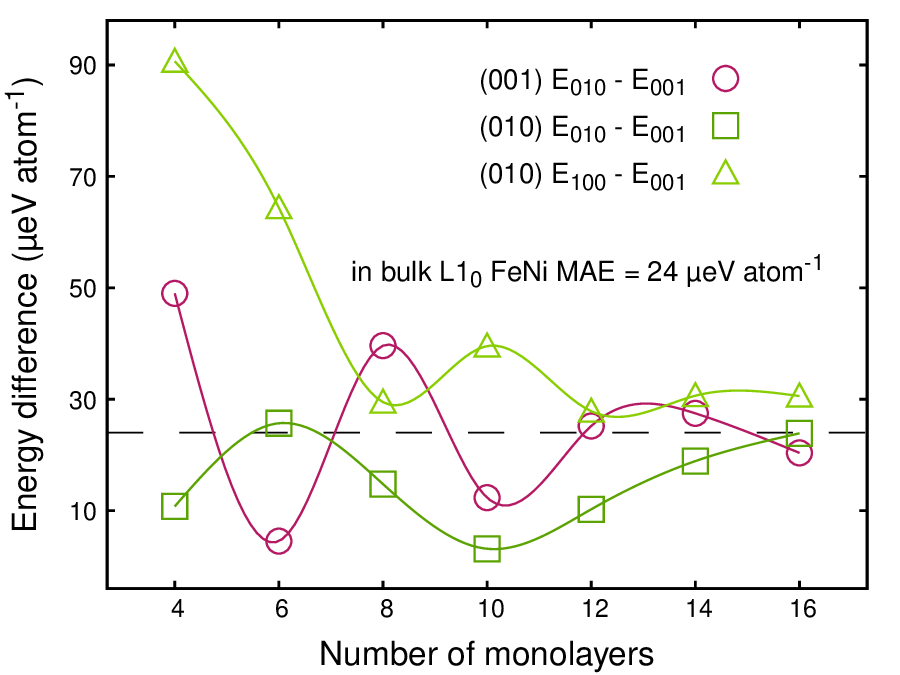}
\caption{\label{fig-de_vs_nml} 
Magnetic anisotropy energies as a function of the number of atomic monolayers of \lozfeni{} (001) and (010) ultrathin films.
The dashed horizontal line indicates the magnetocrystalline anisotropy energy (MAE) calculated for bulk \lozfeni{} (24~$\mu$eV\,atom$^{-1}$)~\cite{werwinski_ab_2017}.
The energies $E_{001}$, $E_{010}$, and $E_{100}$ were determined for the corresponding magnetization directions using the FPLO18 code with the PBE exchange-correlation functional.
}
\end{figure}

%
\begin{figure}[t]
\centering
\includegraphics[clip,width=0.95\columnwidth]{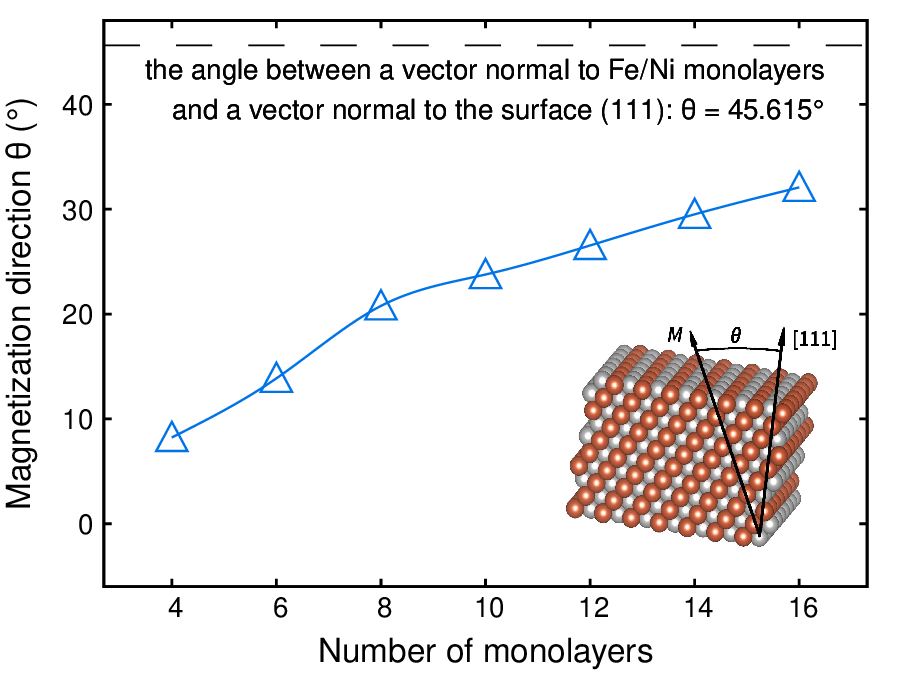}
\caption{\label{fig-dir-vs-nml_111}
Magnetization direction as a function of the number of atomic monolayers of \lozfeni{} (111) thin films. 
The calculations were performed using the FPLO18 code with PBE exchange-correlation potential.
The horizontal dashed line indicates the angle value between the vector [001] normal to the plane of the Fe/Ni atomic monolayers and the vector [111] normal to the film's surface.
}
\end{figure}

%
%
\begin{figure*}[t]
\includegraphics[trim = 0 20 0 0, clip, width=0.8\textwidth]{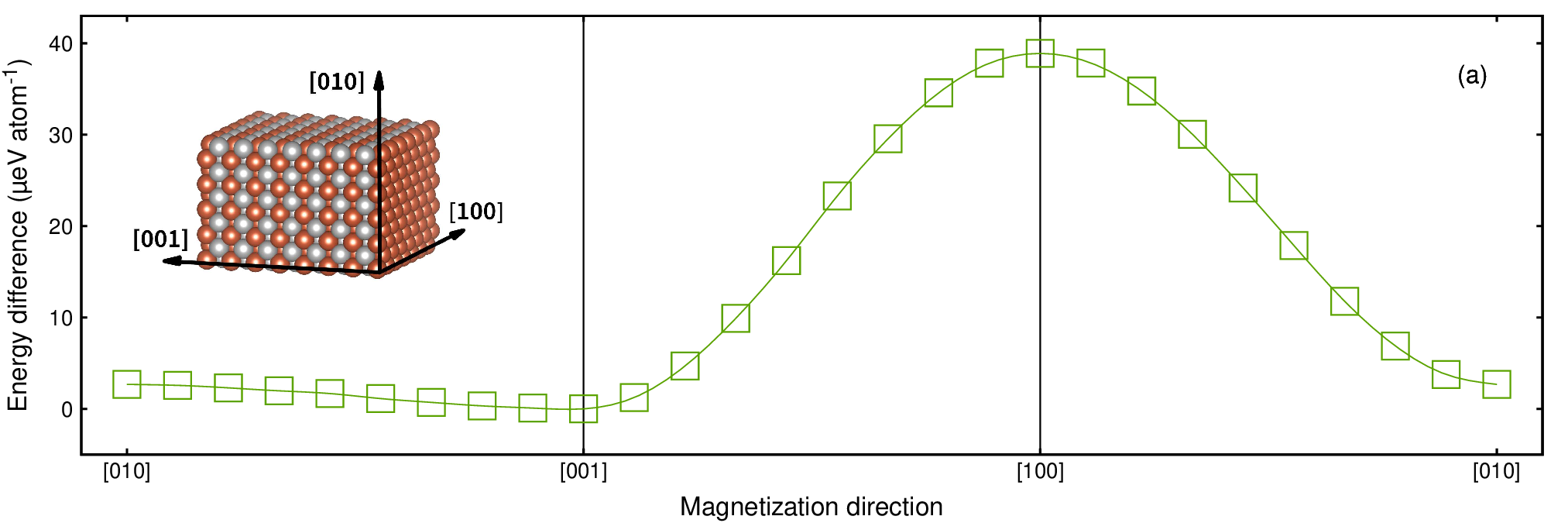}
\includegraphics[trim = 0 0 0 0, clip, width=0.8\textwidth]{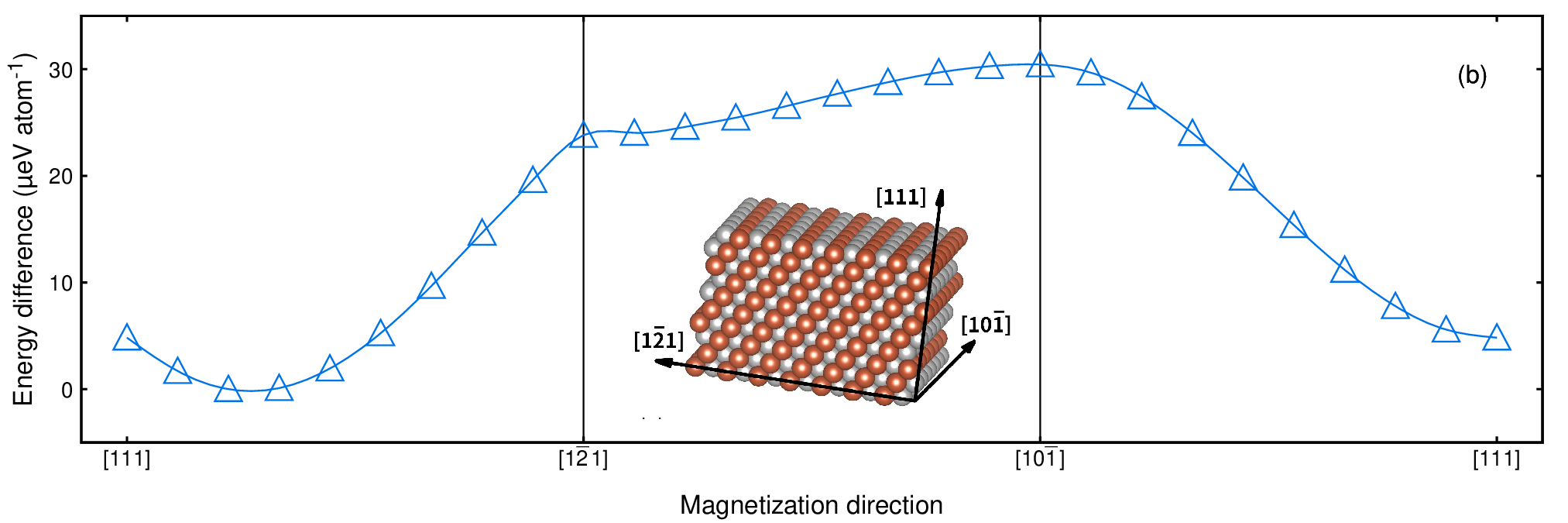}
\caption{\label{fig-e_vs_theta_010_111}
Evolution of magnetic anisotropy energy with magnetization direction for 10-monolayer films of  \lozfeni{} (010) (a) and (111) (b).
The calculations were performed using the FPLO18 code with PBE exchange-correlation functional.
}
\end{figure*}

%
\begin{figure*}[!t]
\hfill
\includegraphics[trim = 75 70 350 75, clip,height = 0.3\textwidth]{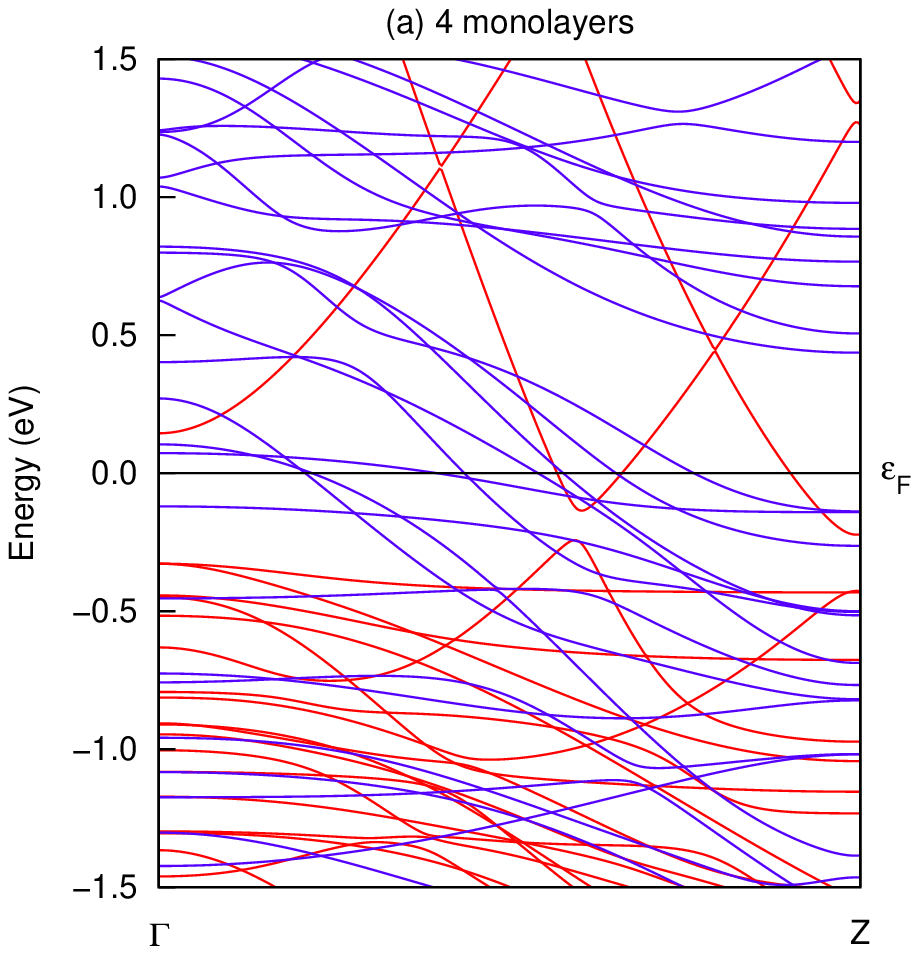}
\includegraphics[trim = 145 70 350 75, clip,height = 0.3\textwidth]{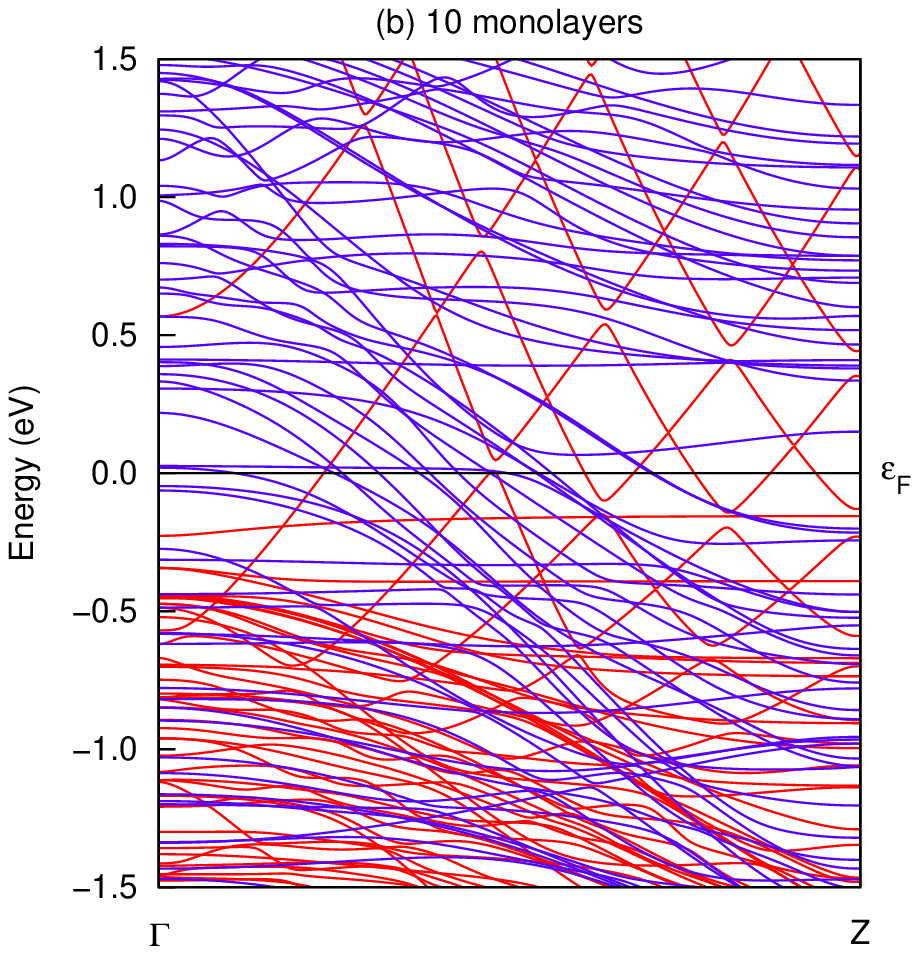}
\includegraphics[trim = 145 70 350 75, clip,height = 0.3\textwidth]{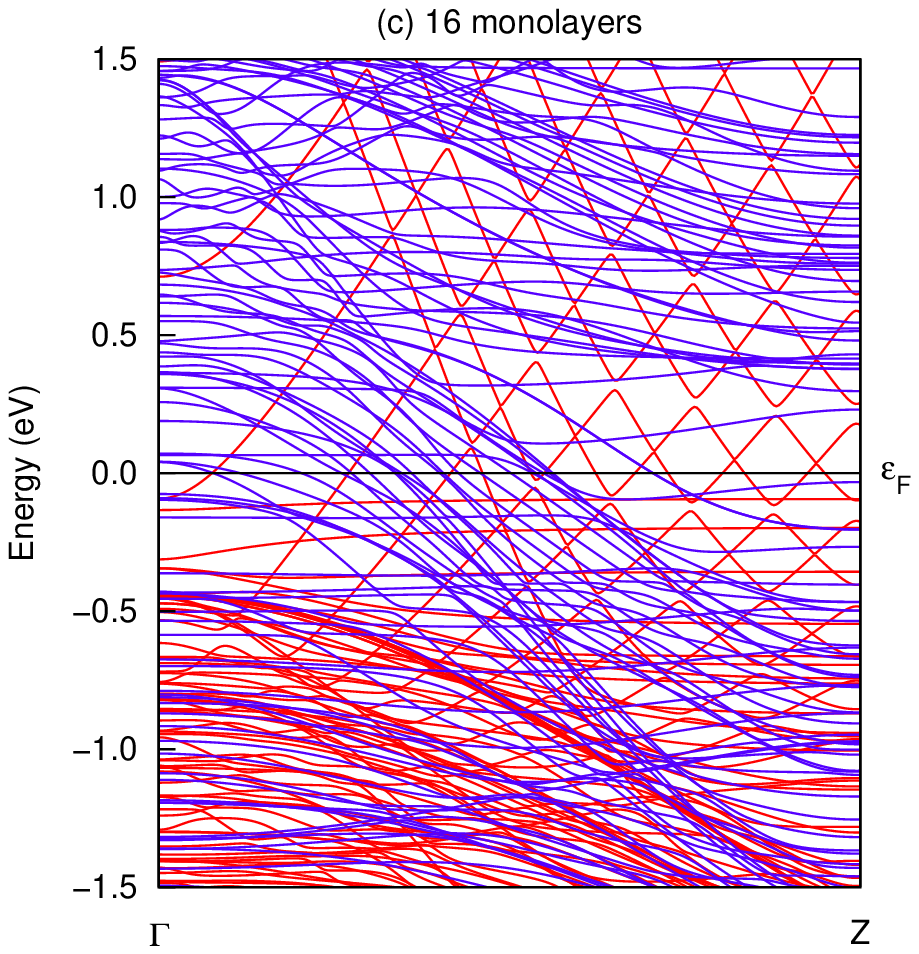}
\includegraphics[trim = 145 70 350 75, clip,height = 0.3\textwidth]{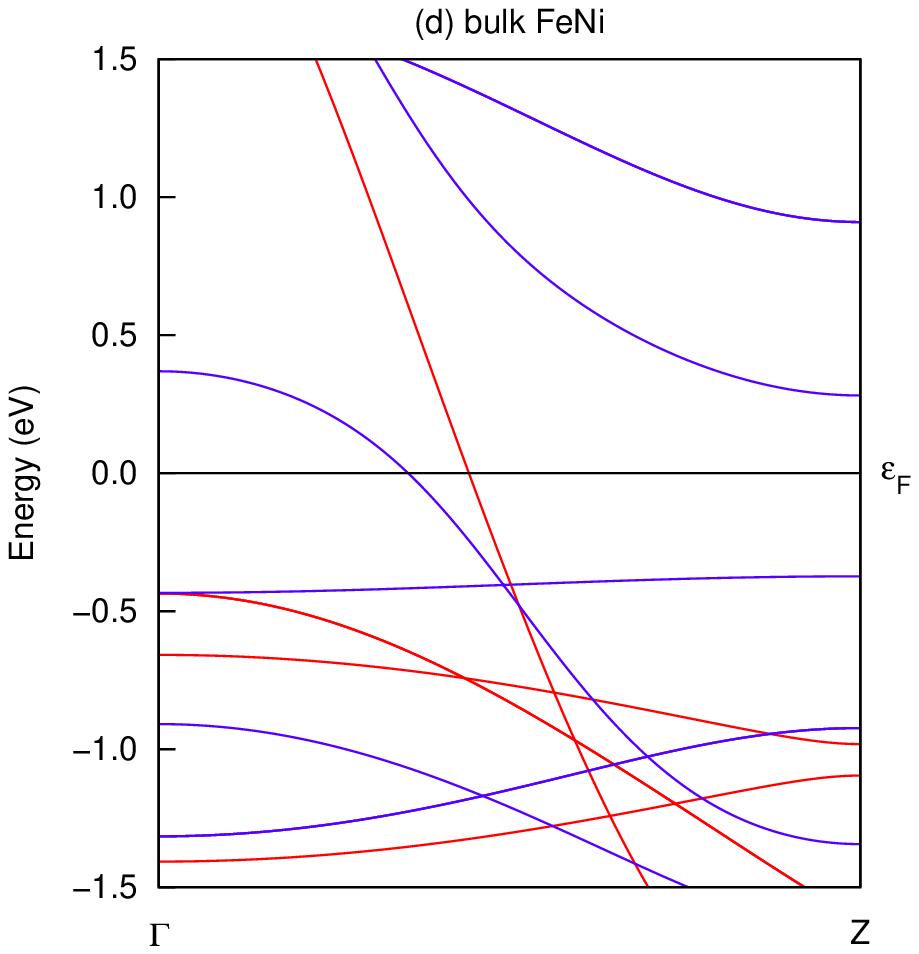}
\hfill
\caption{\label{fig:010_films_bands_gz}
Band structure along $\Gamma$-Z $k$-path calculated for selected \lozfeni{} (010) thin films (a-c) and \lozfeni{} bulk (d).
Red and blue bands denote majority and minority spin channels.
Calculations were performed with the FPLO18 code using the PBE functional and a scalar-relativistic approach.
}
\end{figure*}

%
\begin{figure}[t]
\centering
\includegraphics[trim = 20 10 0 10, clip,width=\columnwidth]{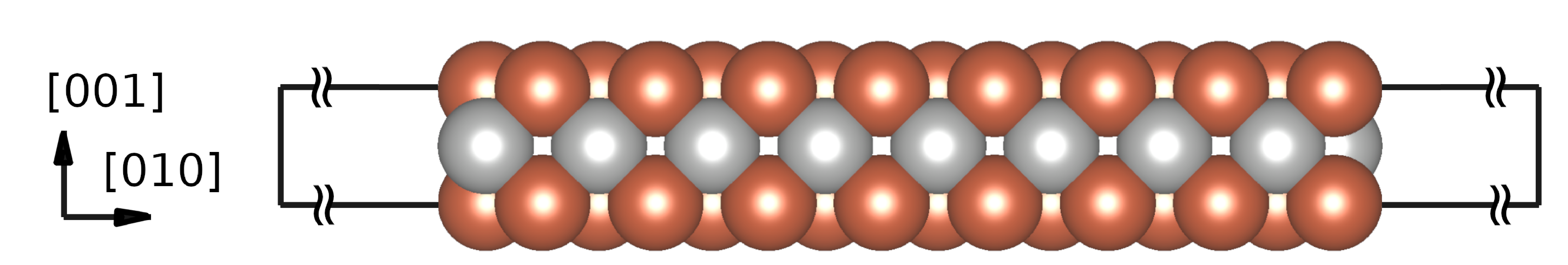}
\vspace{-1mm}

\includegraphics[trim = -10 0 0 0,clip,width=\columnwidth]{feni_010_16mono_ml.eps}
\vspace{-4.5mm}

\includegraphics[clip,width=\columnwidth]{feni_010_16mono_dml.eps}

\caption{\label{fig:010_film_dml}
Orbital magnetic moments at atomic sites along the direction perpendicular to the plane of the film for a 16-monolayer-thick \lozfeni{} (010) film, calculated for magnetization directions [001] and [010] (a).
Difference in orbital magnetic moments calculated between magnetization directions [010] and [001] (b).
Dashed lines indicate the solutions for Fe and Ni calculated for the \lozfeni{} bulk.
Above, a side view of the unit cell of the 16-monolayer FeNi (010) film.
The calculations were performed with the FPLO18 code and the PBE functional.
}
\end{figure}

%
%
The ordered \lozfeni{} was first discovered in neutron bombarded specimens~\cite{pauleve_nouvelle_1962, neel_magnetic_1964, pauleve_magnetization_1968} and later identified in samples cut from meteorites~\cite{petersen_mossbauer_1977,clarke_tetrataeniteordered_1980}.
Studies of natural \loz{} FeNi samples from various meteorites~\cite{
petersen_mossbauer_1977, 
clarke_tetrataeniteordered_1980,
wasilewski_magnetic_1988,
kotsugi_novel_2010, 
kotsugi_structural_2014, 
lewis_inspired_2014, 
poirier_intrinsic_2015}
are being conducted in parallel with research on ordered FeNi thin films produced in laboratories.
%
%
Takanashi, co-author of numerous papers on \lozfeni{} thin films
~\cite{
shima_structure_2007, 
mizuguchi_characterization_2010, 
kojima_l10-ordered_2011, 
mizuguchi_artificial_2011, 
kojima_magnetic_2012,
kotsugi_origin_2013, 
ogiwara_magnetization_2013, 
ohtsuki_nanoscale_2013,
kojima_addition_2014,
mibu_local_2015, 
ueno_structural_2015,
kojima_growth_2016, 
tashiro_fabrication_2018,
ito_epitaxial_2020, 
nishio_fabrication_2021, 
ito_fabrication_2023, 
nishio_uniaxial_2024},
summarized investigations of ordered FeNi thin films in a 2017 topical review~\cite{takanashi_fabrication_2017}.
%
%
The study of Takanashi~\textit{et al.} was complemented by other experimental groups~\cite{chen_laser-induced_2015, 
svalov_study_2015,
frisk_resonant_2016, 
frisk_strain_2017, 
giannopoulos_l10-feni_2018, 
nguyen_ordered_2023}.
Mandal~\etal{} concluded the experimental characteristics of bulk and thin-film \lozfeni{} in a 2023 review article~\cite{mandal_l10_2023}.

%
Measurements of \lozfeni{} films with thickness less than 5~nm~\cite{sakamaki_effect_2013, kojima_feni_2014} give an important context for the calculation results we present in this work.
Kojima~\etal{}~\cite{kojima_feni_2014} have presented the dependence of order parameter, magnetization, and magnetic anisotropy on the number of monolayers.
For thicknesses of about 10 atomic monolayers ($\sim$1.5~nm), low values of both the order parameter and the uniaxial magnetic anisotropy constant were observed~\cite{kojima_feni_2014}.

The magnetic properties of the \lozfeni{} phase are as follows:
Curie temperature of about 830~K~\cite{lewis_inspired_2014},
magnetic moment of about 1.7~$\mu_\mathrm{B}$\,atom$^{-1}$~\cite{frisk_strain_2017},
upper limit of magnetocrystalline anisotropy constants ($K_1$) of 1.3~MJ\,m$^{-3}$~\cite{pauleve_magnetization_1968,lewis_magnete_2014}, and
room-temperature coercivity ($H_\mathrm{c}$) of 95.5~kA\,m$^{-1}$ (1.2~kOe)~\cite{lewis_inspired_2014}.
The transformation of FeNi from an ordered to a disordered phase occurs at a temperature of 593~K (320\st{}C)~\cite{yang_revision_1996,lewis_magnete_2014}.
Whereas obtaining samples with high values of the long-range order parameter ($S$) is difficult~\cite{takanashi_fabrication_2017}.
Kojima \textit{et al.} in research on \lozfeni{} (001) films with a thickness of 50\,$\times$\,(Fe atomic monolayer / Ni atomic monolayer) shown that the uniaxial magnetic anisotropy constant ($K_u$) depends on $S$ and obtained the highest $K_u$ value of 0.7~MJ\,m$^{-3}$ (7.0\,$\times$\,10$^6$~erg\,cm$^{-3}$) for $S = 0.48$~\cite{kojima_magnetic_2012}.
Since this value of $K_u$ is smaller than the shape anisotropy ($2\pi M_\mathrm{s}^2$) estimated for FeNi films as 0.9~MJ\,m$^{-3}$ (9.0\,$\times$\,10$^6$~erg\,cm$^{-3}$), it has not been possible to manufacture FeNi (001) films with a perpendicular magnetic anisotropy so far~\cite{takanashi_fabrication_2017}.
However, Takanashi~\etal{} estimated that increasing the order parameter $S$ above 0.7 should raise the $K_u$ above the shape anisotropy value~\cite{takanashi_fabrication_2017}.
In addition to the lack of complete \loz{} ordering, the films are also not perfectly periodic but contain grains that further negatively affect the magnetic anisotropy.
For the \loz{} phase, it is assumed that three basic nanocrystal variants are present in a single sample, and their main crystallographic axes do not necessarily point in the same direction~\cite{woodgate_revisiting_2023}.

%
The properties of the \lozfeni{} bulk phase were also investigated using density functional theory (DFT)~\cite{
wu_spinorbit_1999,
ravindran_large_2001, 
miura_origin_2013, 
edstrom_electronic_2014, 
lewis_inspired_2014, 
manchanda_transition-metal_2014, 
werwinski_ab_2017, 
tian_density_2019, 
tian_pressure_2020, 
tian_alloying_2021,
tuvshin_fenin_2021,  
si_effect_2022, 
qiao_effect_2023, 
woodgate_revisiting_2023, 
yamashita_finite-temperature_2023}.
The computational method has been used to determine, among others, the magnetocrystalline anisotropy energy (MAE) of a perfect \lozfeni{} crystal~\cite{
wu_spinorbit_1999,
ravindran_large_2001,
edstrom_electronic_2014, 
miura_origin_2013, 
werwinski_ab_2017}, 
to study the effect of chemical disorder~\cite{tian_density_2019, qiao_effect_2023, si_effect_2022, woodgate_revisiting_2023}, and 
to predict how different substitutions affect the MAE~\cite{manchanda_transition-metal_2014, tian_alloying_2021, tuvshin_fenin_2021}.

%
Despite extensive experimental work on \lozfeni{} thin films and numerous computational studies on \lozfeni{} bulk, \lozfeni{} magnetic thin films have not been modeled from first principles.
Only the adsorption of ethylene (C$_2$H$_4$)~\cite{simonetti_study_2010} and 
hydrazine (N$_2$H$_4$)~\cite{he_first-principles_2015} on the \lozfeni{} (111) surface were calculated from DFT.
In this work, continuing our DFT investigation of \loz{} phases~\cite{werwinski_ab_2017, marciniak_dft_2022, marciniak_l10_2023}, we study ultrathin \lozfeni{} films.

%
In this work, the first-principles calculations were made for ultrathin \lozfeni{} films with surfaces (001), (010), and (111).
The study aims to determine the films' magnetic moments and magnetic anisotropies.
Answering the question of how they depend on the thickness and type of surface can help obtain FeNi films with different types of magnetic anisotropy, such as perpendicular, tilted, or fixed in-plane.
Although perpendicular magnetic anisotropy is fairly well known, it is worth briefly discussing the other two types.

%
In 2005, a system with the magnetization direction tilted from the plane of the film was prepared experimentally, and it was shown that deflecting it to 45$^{\circ}$ minimizes the value of the switching magnetic field~\cite{albrecht_magnetic_2005, wang_tilting_2005}.
Since the magnetization direction of \loz{} crystals agrees with the tetragonal axis, a natural candidate for systems with tilted anisotropy are \loz{} (111) films.
Previous experiments on the \lozfept{} (111) films showed that the direction of magnetization is tilted by 33-36° from the plane of the film, i.e., it points roughly in the direction of the tetragonal axis~\cite{zha_pseudo_2009}.
In our computational work on FePt ultrathin films, we also addressed this issue~\cite{marciniak_l10_2023}.
Here, we will determine the magnetization angle for \lozfeni{} (111) films.

%
Alignment of the tetragonal axis leads to perpendicular magnetic anisotropy in films (001) and tilted magnetic anisotropy in films (111).
In films (010), it results in an unconventional type of in-plane anisotropy with a preferred direction in the film plane.
We already postulated that such a uniaxial configuration could be called \textit{fixed in-plane} magnetic anisotropy~\cite{marciniak_l10_2023}, as opposed to conventional \textit{in-plane}.
20~nm \lozfeni{} (110) films with a tetragonal axis in the plane of the film were experimentally obtained in 2023 and showed order parameter $S$ of 0.6 and magnetic anisotropy constant of 0.55~\MJ{}~\cite{ito_fabrication_2023}.
Experimental work on isostructural \lozfept{} (010) films are also known~\cite{hsu_situ_2000,shima_preparation_2002,ohtake_l10_2012,sepehri-amin_microstructure_2017,wu_atomic-scale_2022}.
In this work, we will determine the magnetic easy axis of \lozfeni{} (010) films.

\section{Calculations' details}

%
We prepared models of \lozfeni{} (001), (010), and (111) thin films in the thickness range from four to sixteen atomic monolayers, see  Fig.~\ref{fig-structures}.
To keep the concentration of Fe and Ni equal, we considered only films with an even number of monolayers.
As a starting point for modeling thin films, we used a unit cell of the \lozfeni{} bulk phase (s.g. $P$4$/mmm$, $a$'~=~3.56/$\sqrt{2} \sim 2.52$ and $c$~=~3.58~\AA{})~\cite{werwinski_ab_2017}.
We added a minimum of 30~\AA{} vacuum to all unit cells of the thin films in the direction perpendicular to the film plane.

%
For the prepared models, calculations were performed using the full-potential local-orbital (FPLO18.00-52) code~\cite{koepernik_full-potential_1999, opahle_full-potential_1999} within the density functional theory (DFT).
The exchange-correlation potential in the Perdew-Burke-Ernzernhof (PBE) form was used~\cite{perdew_generalized_1996}. 
%
%
The atomic positions of the films were optimized using forces in a spin-polarized approach in 20\,$\times$\,20\,$\times$\,1 \textbf{k}-point mesh, with a convergence criterion of 10$^{-3}$\,eV\,\AA{}$^{-1}$.
Table~\ref{tab-wyckoff-positions} shows space groups, calculated Wyckoff positions, and unit cell parameters of the example 16-monolayer films.
The prepared (001) and (010) films have thicknesses from 0.5 to 2.7~nm, while the (111) films span from 0.6 to 3.1~nm (all values determined as the distance between the atomic positions of the surface sites).

%
The \textbf{k}-point mesh was considerably densified in the next step to determine the accurate magnetocrystalline anisotropy energy (MAE) values.
For (001) and (010) films, 
a \textbf{k}-point mesh of 60\,$\times$\,60\,$\times$\,5 was assumed, 
and for (111) films, a \textbf{k}-point mesh of 60\,$\times$\,30\,$\times$\,5.
For Brillouin zone integration, the tetrahedron method was chosen.
For scalar-relativistic calculations, a 10$^{-6}$ density convergence criterion was used.
After self-consistent scalar-relativistic calculations, one iteration of fully relativistic calculations was performed for the selected quantization axes.
Such fully relativistic solutions were used to determine magnetic anisotropy energies and energy dependencies as a function of the quantization axis (magnetization) direction.
As demonstrated~\cite{marciniak_structural_2023}, this approach leads to sufficiently accurate results, significantly reducing computation time.
%
%
Calculations of the energy dependence of the magnetization direction were performed for 10-monolayer films (010) and (111) using unit cells with space group \textit{P}1. 
For (111) films, having established that the energy minimum is in the (100) plane, calculations were carried for directions varying in this plane every 2$^{\circ}$.
The direction of easy magnetization was then determined using third-degree polynomial fitting.

%
The unique implementation of the fixed spin moment method in a fully relativistic approach of the FPLO code allowed us to determine the dependence of the magnetocrystalline anisotropy energy on the spin magnetic moment. 
We performed these calculations with the PBE exchange-correlation functional in the generalized gradient approximation (GGA), as well as with the Perdew-Wang (PW92) functional in the local density approximation (LDA)~\cite{perdew_accurate_1992}.

We calculated the $k$-point resolved contributions to MAE using the magnetic force theorem~\cite{liechtenstein_local_1987,wang_validity_1996,wu_spinorbit_1999} defined by equation:
\begin{multline}\label{eq:mag_force}
\mathrm{MAE} = E(\theta = 90^{\circ}) -  E(\theta = 0^{\circ}) =\\
= \sum_{\textrm{occ'}} \epsilon_{i}(\theta = 90^{\circ}) -  \sum_{\textrm{occ''}} \epsilon_{i}(\theta = 0^{\circ}), 
\end{multline}
where
$E(\theta)$ is the total energy for a specific direction;
$\theta$ is the angle between the direction of magnetization and the $c$ axis;
$\epsilon_{i}$ is the band energy of state $i$.

Because the implementation of the DFT used in this work is based on the linear combination of atomic orbitals, it was possible to perform population analysis using Mulliken's method~\cite{mulliken_electronic_1955}.
We used the VESTA code to prepare drawings of the structures~\cite{momma_vesta_2008}.
In our previous article on ultrathin films of \loz{} FePt, we discussed the limitations of the models adopted and the approximations used~\cite{marciniak_l10_2023}.

\section{Results and discussion}

Before discussing the calculated results for the films, we will present an analysis of the dependence of the magnetic anisotropy energy on the value of the magnetic moment and the form of the exchange-correlation potential for bulk \lozfeni{}.
This allows us to place in a context the magnetic anisotropy results for ultrathin FeNi films, which in the limit of large thicknesses converge to the values for the bulk.

\subsection{Magnetocrystalline anisotropy of bulk \lozfeni{}}
%
%
For bulk \lozfeni{}, the MAE of 0.34~\MJ{} (24~$\mu$eV\,atom$^{-1}$) determined from the GGA is much lower than the maximum determined values of experimental magnetic anisotropy constants of up to 1.3~\MJ{}~\cite{lewis_magnete_2014}.
The discrepancy may be due to the temperature for which the DFT calculations are done (by definition, 0~K) or the limitations of the GGA.
Furthermore, in our work on bulk \lozfept{}, we pointed out that the values of MAE and magnetic moment are correlated~\cite{marciniak_dft_2022}, and the latter depends on the choice of the exchange-correlation potential.
%
%
%
To show this correlation also for bulk \lozfeni{}, we performed calculations of the dependence of MAE on a fixed spin moment for LDA and GGA functional.
Figure~\ref{fig-bulk_MAE_vs_fsm} shows that for \lozfeni{} the equilibrium values of magnetic moments and MAEs obtained in the PW92 and PBE approximations differ only slightly.
Moreover, the MAE value depends on the magnetic moment, and in both approximations, it has a maximum above 1.5~\MJ{} for a reduced magnetic moment of about 1.35~\muB\,atom$^{-1}$.
The presented relation of MAE and magnetic moment may partially explain the discrepancy between the experimentally determined $K_u$ values above 1~\MJ{} at room temperature and the calculated MAE values below 0.5~\MJ{} at 0~K.
Namely, $K_u$ can also indirectly depend on temperature via temperature-mediated reduction of magnetization.
Such interpretation is supported by calculations of the MAE dependence on temperature for \loz{}~Fe$_{0.56}$Ni$_{0.44}$, showing lowered MAE at low temperatures and MAE maximum near room temperature~\cite{woodgate_revisiting_2023}.
Furthermore, $K_u$ measurements of \lozfeni{} films performed at 35~K reveal much lower $K_u$ values (<~0.2~\MJ{}) than typical results obtained at room temperature~\cite{frisk_resonant_2016}.

%
To understand the dependence of the MAE on the spin magnetic moment, we look at 
the density of states and 
the band structure of the bulk \lozfeni{}.
Figure~\ref{fig:bulk_dos_bands_bzone} presents the equilibrium solutions (without fixed spin moment) of these properties.
The energy range in Fig.~\ref{fig:bulk_dos_bands_bzone} covers most of the valence band.
The majority spin channel (red) is mostly occupied, while the minority spin channel (blue) has empty states above the Fermi level.

\subsection{\lozfeni{} ultrathin films}

In the following subsection, we discuss magnetic moments, magnetization directions, and magnetic anisotropy energies of considered films.

\subsubsection{Energetic stability, magnetic moments, and excess electrons of ultrathin films}
The total energy of the films decreases with thickness, tending asymptotically to the bulk value, see Fig.~\ref{fig-energy_vs_nml}.
Over the considered range, the most energetically stable films are (111), then (010), and last (001),
with the energy of the (111) surface being much lower than the other two.
Previously, the same order was determined for the surfaces of \loz{} FePt and \loz{} CoPt films~\cite{dannenberg_surface_2009}.
This result identifies a suboptimal energy condition for (001) surface formation

%

%
Figure~\ref{fig-m_vs_position} shows excess electrons on Fe and Ni atoms along the thickness of films, which were calculated using Mulliken's approach~\cite{mulliken_electronic_1955}.
The deviations in excess electrons on FeNi surfaces (ranging from about -0.02 to 0.15 on Fe sites) are much smaller than those calculated previously for FePt surfaces (ranging from about -1.0 to 1.0 on Fe sites)~\cite{marciniak_l10_2023}.
This indicates higher stability of FeNi surfaces in comparison to L1$_0$ FePt ultrathin films.
As we look closer at this plot, the excess electron (hole) values are close to those for bulk FeNi in the central region of the films, while for about three surface monolayers, they show large deviations from the bulk value.
We observed similar behavior in the results of additional calculations performed without spin polarization, confirming the primary role of charge transfer in the formation of perturbations in the near-surface monolayers.

Figure~\ref{fig-m_vs_position} also presents the spin and orbital magnetic moments on individual atoms along film thickness.
We can see a similar increase in magnetic moments on about three surface monolayers and values close to the bulk value in the central region~\cite{werwinski_ab_2017}.
The highest values of magnetic moments are on the surface (010).
Magnetic moments on the surfaces are elevated throughout the thickness range considered; the central region vanishes only below about six monolayers.
Unlike the (010) and (111), the model of (001) film has asymmetrical \textit{top} and \textit{bottom} surfaces (one side ends in an Fe monolayer and the other in Ni).
Hence, the observed asymmetry in the (001) results in Figure~\ref{fig-m_vs_position}.

\subsubsection{Magnetic anisotropy of ultrathin films}

%
For ultrathin films of \lozfeni{}, we observed an increase in the average magnetic moment only to about 5\% over the bulk value.
However, the magnetic anisotropy of thin films differs much more.
In the case of magnetic anisotropy, we are interested in both the value of the magnetic anisotropy energy and the direction of magnetization relative to the plane of the film (magnetic easy axis).
We expect that since the films under consideration have an \loz{} structure with a unique tetragonal axis, the direction of magnetization will be along or close to this axis.
This means that perpendicular, in-plane, and tilted magnetic anisotropy should be observed for the films (001), (010), and (111), respectively.

%
We will begin our discussion of the magnetic anisotropy of  \lozfeni{} films with the most studied (001) films. 
As in the previous experiments
~\cite{
shima_structure_2007, 
mizuguchi_characterization_2010, 
kojima_l10-ordered_2011, 
mizuguchi_artificial_2011, 
kojima_magnetic_2012,
kotsugi_origin_2013, 
ogiwara_magnetization_2013, 
ohtsuki_nanoscale_2013,
kojima_addition_2014,
mibu_local_2015, 
ueno_structural_2015,
kojima_growth_2016, 
tashiro_fabrication_2018,
ito_epitaxial_2020, 
nishio_fabrication_2021, 
ito_fabrication_2023, 
nishio_uniaxial_2024,
takanashi_fabrication_2017},
we also expect for them the perpendicular magnetic anisotropy. 
The relation of total energy and magnetic moment to the number of monolayers suggests that we should also expect a dependence on magnetic anisotropy energy that converges asymptotically to the bulk value.
However, as we can see in Fig.~\ref{fig-de_vs_nml}, although the anisotropy energy of the (001) films converges to the bulk value, this is accompanied by pronounced oscillations.
This type of MAE behavior has been observed in several previous calculations for ultrathin films~\cite{szunyogh_oscillatory_1997,zhang_electric-field_2009,blanco-rey_large_2021,chang_voltage-controlled_2021, cinal_magnetic_2022, marciniak_l10_2023}.
Oscillations of uniaxial magnetic anisotropy have also been observed experimentally for ultrathin bcc Fe and fcc Co films with periods of 5.9 and 2.3 atomic monolayers, respectively~\cite{przybylski_oscillatory_2012}.
They are due to quantum well states situated around the Fermi level~\cite{przybylski_oscillatory_2012} and come from the quantum size effect appearing in small enough systems.
The positive values of the magnetic anisotropy energies for films (001), defined as energy difference $E_{010} - E_{001}$, indicate a preference for orientation of the magnetization direction perpendicular to the plane of the (001) film.
However, the result obtained does not clearly identify perpendicular anisotropy, since for this, the uniaxial anisotropy energy must exceed the energy of shape anisotropy ($2 \pi M_s^2$~$\sim$~0.9~\MJ{}~\cite{kojima_feni_2014}).
This criterion is not met in the case of our computational results.
Nevertheless, uniaxial anisotropy is still expected to overcome the shape anisotropy in experiments at room temperature, where the determined $K_u$ value is up to three times higher than the GGA value calculated here.

For films (010) with Fe/Ni monolayers aligned perpendicular to the plane of the film (\textit{bookshelf} arrangement), we expect an easy magnetization axis in the direction of the crystallographic tetragonal axis, that is, in the plane of the film.
Indeed, this magnetization direction is the most energetically stable in this case, see Figs.~\ref{fig-de_vs_nml} and \ref{fig-e_vs_theta_010_111}.
However, since magnetization has a clear preference in the direction of the tetragonal axis [001], this is not a typical case of in-plane magnetic anisotropy.
The corresponding phenomenon observed in FePt (010) films we proposed to call \textit{fixed in-plane} magnetic anisotropy~\cite{marciniak_l10_2023}.
As in the case of the (001) films discussed above, we also observe oscillations in the magnetic anisotropy energy due to the quantum size effect.
Furthermore, as mentioned, \textit{bookshelf}-type FeNi films with a (110) surface were obtained in 2023 experimentally, but the dependence of the magnetization anisotropy in the plane of the films was not investigated~\cite{ito_fabrication_2023}.
It is then not clear how shape anisotropy will affect uniaxial anisotropy in this case.

%
To give an idea of the physical origins of magnetic anisotropy in (010) films, below we will present an analysis of the difference in orbital moments based on Bruno's formula for an example film with a thickness of 16 monolayers, and present images of the band structure for (010) films of different thicknesses.

%
Figure~\ref{fig:010_films_bands_gz} presents the evolution of the band structure (along the $\Gamma$-Z $k$-path) with the thickness of the (010) films.
Both similarities and differences between film and bulk solutions are noticeable.
The number of bands increases with the thickness of the film increases, which is due to an increase in the number of non-equivalent sites in the system.
In the limit of larger thicknesses, the solution includes contributions from the central part of the film (similar to the solution for bulk), contributions from atoms in the near-surface layers, and also quantum well states~\cite{przybylski_oscillatory_2012}.
The change in band structure with film thickness consequently leads to oscillation of the MAE with film thickness.
In turn, in the limit of large thicknesses, the film's MAE converges to the value for bulk material and the surface contributions are negligible.

%
MAE contributions can be analyzed from the perspective of real space, using Bruno's formula~\cite{bruno_tight-binding_1989} linking the magnetic anisotropy energy to the anisotropy in orbital magnetic moments:
\begin{equation}
\mathrm{MAE} \sim \pm \frac{\xi}{4} \Delta m_L,
\end{equation}
where $\xi$ is the spin-orbit constant and
$\Delta m_L$ is the difference in the orbital magnetic moment defined as:
\begin{equation}
\Delta m_L = m_L^{||} - m_L^{\perp},
\end{equation}
where $m_L^{||}$ and $m_L^{\perp}$ are the orbital magnetic moments calculated for two perpendicular quantization axes.
For bulk \lozfeni{} the calculated orbital magnetic moments 
for Fe are 0.054~$\mu_\mathrm{B}$ (0.048~$\mu_\mathrm{B}$) 
in [001] ([110]) magnetization direction and 
for Ni they are 0.038~$\mu_\mathrm{B}$ (0.041~$\mu_\mathrm{B}$) 
in [001] ([110]) direction.
The calculated differences in the orbital magnetic moment are $\Delta m_L$(Fe) = -0.0060~$\mu_\mathrm{B}$ and $\Delta m_L$(Ni) = 0.0022~$\mu_\mathrm{B}$,
leading to a total $\Delta m_L$ equal to -0.0038~$\mu_\mathrm{B}$\,f.u.$^{-1}$.
The normalized contributions of Fe and Ni to $\Delta m_L$ are 1.58 and -0.58, respectively. 
The above values are consistent with the sublattice-resolved MAE calculated by Ke for \lozfeni{} using perturbation theory in the tight-binding approach, resulting in contributions of 1.57 and -0.57 for Fe and Ni, respectively (for MAE calculated as 33~$\mu$eV\,atom$^{-1}$)~\cite{ke_intersublattice_2019}.
In Fig.~\ref{fig:010_film_dml}, we show the orbital magnetic moments and their differences calculated for the two magnetization directions for an example 16-monolayer \lozfeni{} (010) film.
In several near-surface layers, we see the largest deviations of orbital magnetic moments and differences of orbital magnetic moments from the values in the central part of the film.
At the same time, the values of orbital magnetic moment differences in the central part are close to the values for bulk \lozfeni{}.

%
Finally, we will discuss the magnetic anisotropy of (111) films.
We do not present for them the dependence of the anisotropy energy on the number of monolayers, 
because the direction of magnetization changes with film thickness.
For \lozfeni{} (111) films, the angle of the normal to the alternating Fe and Ni monolayers (the tetragonal axis [001]) to the out-of-plane direction [111] is 45.615$^o$.
This value also specifies the easy magnetization axis within the limits of thick films.
For ultrathin films, however, we expect the magnetic easy axis to deviate from the tetragonal axis [001].
The determined direction of the magnetic easy axis for the (111) film with a thickness of 10 atomic monolayers (1.85~nm) is deviated from the [111] direction by 23.8$^{\circ}$, see Fig.~\ref{fig-e_vs_theta_010_111}.
Furthermore, Fig.~\ref{fig-dir-vs-nml_111} shows that the direction of the easy axis indeed converges asymptotically to the hypothetical value of 45.615$^o$ for the thick (111) films.
Whereas, for the thinnest films, the easy axis approaches the out-of-plane direction [111].

Although it might be counterintuitive, perhaps it is the ultrathin \lozfeni{} films (111), rather than films (001), that will allow for the practical realization of perpendicular magnetic anisotropy?
In their favor is the highest energy stability among the three surfaces considered, and the positive results of recent experiments confirming the high values of both the order parameter $S$ and the anisotropy constant $K_u$ (1.15~\MJ{})~\cite{nguyen_ordered_2023}.
On the other hand, with the uniaxial anisotropy constant tuned in such a way that it only slightly exceeds the shape anisotropy and the direction of the easy axis at about 45$^{\circ}$ to the surface, the \lozfeni{} (111) thin films can become excellent systems for devices with switching magnetic field~\cite{albrecht_magnetic_2005, wang_tilting_2005}.
%
%
There are, however, limitations related to the demagnetization effect, which may prevent the magnetization direction in the films from being set in the easy axis predicted from DFT~\cite{kovacs_micromagnetic_2017,skomski_magnetic_2016}.
Skomski and Coey pointed out that for a magnet of any shape (including film) to resist demagnetization, the anisotropy field must exceed the demagnetizing field, for which the condition is $K_1 > 1/2 \mu_0 M_s^2$~\cite{skomski_magnetic_2016}.
In case of our models calculated at 0~K, with saturated total magnetic moment of bulk \lozfeni{} equal to 3.34~$\mu_B$\,f.u.$^{-1}$ ($M_s = 1.36$~MA\,m$^{-1}$), we estimate that to resist demagnetization the $K_1$ (MAE) should exceed 1.16~\MJ{} (about 80~$\mu$eV\,atom$^{-1}$).
Unfortunately, this condition is not fulfilled, as our 0~K PBE value of MAE is only 0.34~\MJ{} (24~$\mu$eV\,atom$^{-1}$) for bulk \lozfeni{}, and stay around it also for considered thin films.
This indicates that the demagnetization effect will not allow easy magnetization axes of the films to be fixed in the directions predicted from DFT calculations.
However, the condition $K_1 > 1/2 \mu_0 M_s^2$ might be fulfilled in practice.
The measured magnetization of the FeNi films is lower than that predicted from PBE, leading to a reduction in the $K_1$ value necessary to meet the demagnetization resistance condition below 1.0~\MJ{}.
At the same time, some measured MAE values, such as for FeNi (111) films, are as high as 1.15~\MJ{}~\cite{nguyen_ordered_2023}.

\section{Summary and conclusions}
The computational results presented here extend previous research on \lozfeni{} films to include the thinnest films with thickness from 0.5 to 3~nm. 
Among the films with (001), (010) and (111) surfaces, the latter is the most energetically preferable. 
Reducing the thickness of ultrathin films raises the magnetic moment by about 5\%. 
Furthermore, in films (001) and (010), the easy magnetization direction follows the direction of the crystallographic tetragonal axis and is perpendicular to and in the plane of the film, respectively.
At the same time, the alignment of the easy axis in the (010) film plane is unusual in that it is strongly anisotropic in the plane, clearly preferring the direction of the tetragonal axis [001].
For thicker (111) films, the easy magnetization direction is consistent with the tetragonal axis direction [001] (at the angle of about 45$^{\circ}$ to the normal to the plane).
However, for ultrathin (111) films, it deviates from [001] towards [111] direction (normal to the plane).
By this, the magnetic anisotropy of the thinnest (111) films resembles perpendicular anisotropy.

The unique anisotropic properties of \lozfeni{} films may find application in new spintronic devices, as in the case of \lozfept{} films.
However, the uniaxial magnetic anisotropy for the \lozfeni{} films is an order of magnitude lower than for the \lozfept{} films and competes with the shape anisotropy.
Still, the lower magnetic anisotropy constant may be preferable for some applications, and the significant price difference between nickel and platinum may also favor FeNi films.

\section*{Acknowledgments}
We acknowledge the financial support of the National Science Centre Poland under the decision DEC-2018/30/E/ST3/00267.
Part of the computations were performed using resources provided by the Poznan Supercomputing and Networking Center (PSNC).
We thank P. Leśniak and D. Depcik for compiling the scientific software and administration of the computing cluster at the Institute of Molecular Physics, Polish Academy of Sciences.
We also thank Z. Śniadecki and J. Snarski-Adamski for reading the manuscript and providing valuable comments.

\end{sloppypar}

\bibliography{feni_films}

\begin{thebibliography}{10}
\expandafter\ifx\csname url\endcsname\relax
  \def\url#1{\texttt{#1}}\fi
\expandafter\ifx\csname urlprefix\endcsname\relax\def\urlprefix{URL }\fi
\expandafter\ifx\csname href\endcsname\relax
  \def\href#1#2{#2} \def\path#1{#1}\fi

\bibitem{werwinski_ab_2017}
M.~Werwi{\'n}ski, W.~Marciniak, Ab initio study of magnetocrystalline
  anisotropy, magnetostriction, and {{Fermi}} surface of
  {{L1}}{\textsubscript{0}} {{FeNi}} (tetrataenite), J. Phys. Appl. Phys.
  50~(49) (2017) 495008.
\newblock \href {https://doi.org/10.1088/1361-6463/aa958a}
  {\path{doi:10.1088/1361-6463/aa958a}}.

\bibitem{pauleve_nouvelle_1962}
J.~Paulev{\'e}, D.~Dautreppe, J.~Laugier, L.~N{\'e}el, {Une nouvelle transition
  ordre-d{\'e}sordre dans Fe-Ni (50-50)}, J. Phys. Radium 23~(10) (1962)
  841--843.
\newblock \href {https://doi.org/10.1051/jphysrad:019620023010084100}
  {\path{doi:10.1051/jphysrad:019620023010084100}}.

\bibitem{neel_magnetic_1964}
L.~N{\'e}el, J.~Pauleve, R.~Pauthenet, J.~Laugier, D.~Dautreppe, Magnetic
  {{Properties}} of an {{Iron-Nickel Single Crystal Ordered}} by {{Neutron
  Bombardment}}, J. Appl. Phys. 35~(3) (1964) 873--876.
\newblock \href {https://doi.org/10.1063/1.1713516}
  {\path{doi:10.1063/1.1713516}}.

\bibitem{pauleve_magnetization_1968}
J.~Paulev{\'e}, A.~Chamberod, K.~Krebs, A.~Bourret, Magnetization {{Curves}} of
  {{Fe}}--{{Ni}} (50--50) {{Single Crystals Ordered}} by {{Neutron
  Irradiation}} with an {{Applied Magnetic Field}}, J. Appl. Phys. 39~(2)
  (1968) 989--990.
\newblock \href {https://doi.org/10.1063/1.1656361}
  {\path{doi:10.1063/1.1656361}}.

\bibitem{petersen_mossbauer_1977}
J.~F. Petersen, M.~Aydin, J.~M. Knudsen, M{\"o}ssbauer spectroscopy of an
  ordered phase (superstructure) of {{FeNi}} in an iron meteorite, Phys. Lett.
  A 62~(3) (1977) 192--194.
\newblock \href {https://doi.org/10.1016/0375-9601(77)90023-8}
  {\path{doi:10.1016/0375-9601(77)90023-8}}.

\bibitem{clarke_tetrataeniteordered_1980}
R.~S. Clarke, E.~R.~D. Scott, Tetrataenite---ordered {{FeNi}}, a new mineral in
  meteorites, Am. Mineral. 65~(7-8) (1980) 624--630.

\bibitem{wasilewski_magnetic_1988}
P.~Wasilewski, Magnetic characterization of the new magnetic mineral
  tetrataenite and its contrast with isochemical taenite, Phys. Earth Planet.
  Inter. 52~(1) (1988) 150--158.
\newblock \href {https://doi.org/10.1016/0031-9201(88)90063-5}
  {\path{doi:10.1016/0031-9201(88)90063-5}}.

\bibitem{kotsugi_novel_2010}
M.~Kotsugi, C.~Mitsumata, H.~Maruyama, T.~Wakita, T.~Taniuchi, K.~Ono,
  M.~Suzuki, N.~Kawamura, N.~Ishimatsu, M.~Oshima, Y.~Watanabe, M.~Taniguchi,
  Novel {{Magnetic Domain Structure}} in {{Iron Meteorite Induced}} by the
  {{Presence}} of {{L1}}{\textsubscript{0}}-{{FeNi}}, Appl. Phys. Express 3~(1)
  (2010) 013001.
\newblock \href {https://doi.org/10.1143/APEX.3.013001}
  {\path{doi:10.1143/APEX.3.013001}}.

\bibitem{kotsugi_structural_2014}
M.~Kotsugi, H.~Maruyama, N.~Ishimatsu, N.~Kawamura, M.~Suzuki, M.~Mizumaki,
  K.~Osaka, T.~Matsumoto, T.~Ohkochi, T.~Ohtsuki, T.~Kojima, M.~Mizuguchi,
  K.~Takanashi, Y.~Watanabe, Structural, magnetic and electronic state
  characterization of {{L1}}{\textsubscript{0}}-type ordered {{FeNi}} alloy
  extracted from a natural meteorite, J. Phys. Condens. Matter 26~(6) (2014)
  064206.
\newblock \href {https://doi.org/10.1088/0953-8984/26/6/064206}
  {\path{doi:10.1088/0953-8984/26/6/064206}}.

\bibitem{lewis_inspired_2014}
L.~H. Lewis, A.~Mubarok, E.~Poirier, N.~Bordeaux, P.~Manchanda, A.~Kashyap,
  R.~Skomski, J.~Goldstein, F.~E. Pinkerton, R.~K. Mishra, R.~C. Kubic~Jr,
  K.~Barmak, Inspired by nature: Investigating tetrataenite for permanent
  magnet applications, J. Phys. Condens. Matter 26~(6) (2014) 064213.
\newblock \href {https://doi.org/10.1088/0953-8984/26/6/064213}
  {\path{doi:10.1088/0953-8984/26/6/064213}}.

\bibitem{poirier_intrinsic_2015}
E.~Poirier, F.~E. Pinkerton, R.~Kubic, R.~K. Mishra, N.~Bordeaux, A.~Mubarok,
  L.~H. Lewis, J.~I. Goldstein, R.~Skomski, K.~Barmak, Intrinsic magnetic
  properties of {{L1}}{\textsubscript{0}} {{FeNi}} obtained from meteorite
  {{NWA}} 6259, J. Appl. Phys. 117~(17) (2015) 17E318.
\newblock \href {https://doi.org/10.1063/1.4916190}
  {\path{doi:10.1063/1.4916190}}.

\bibitem{shima_structure_2007}
T.~Shima, M.~Okamura, S.~Mitani, K.~Takanashi, Structure and magnetic
  properties for {{L1}}{\textsubscript{0}}-ordered {{FeNi}} films prepared by
  alternate monatomic layer deposition, J. Magn. Magn. Mater. 310~(2, Part 3)
  (2007) 2213--2214.
\newblock \href {https://doi.org/10.1016/j.jmmm.2006.10.799}
  {\path{doi:10.1016/j.jmmm.2006.10.799}}.

\bibitem{mizuguchi_characterization_2010}
M.~Mizuguchi, S.~Sekiya, K.~Takanashi, Characterization of {{Cu}} buffer layers
  for growth of {{L1}}{\textsubscript{0}}-{{FeNi}} thin films, J. Appl. Phys.
  107~(9) (2010) 09A716.
\newblock \href {https://doi.org/10.1063/1.3337649}
  {\path{doi:10.1063/1.3337649}}.

\bibitem{kojima_l10-ordered_2011}
T.~Kojima, M.~Mizuguchi, K.~Takanashi, L1{\textsubscript{0}}-ordered {{FeNi}}
  film grown on {{Cu-Ni}} binary buffer layer, J. Phys. Conf. Ser. 266 (2011)
  012119.
\newblock \href {https://doi.org/10.1088/1742-6596/266/1/012119}
  {\path{doi:10.1088/1742-6596/266/1/012119}}.

\bibitem{mizuguchi_artificial_2011}
M.~Mizuguchi, T.~Kojima, M.~Kotsugi, T.~Koganezawa, K.~Osaka, K.~Takanashi,
  Artificial {{Fabrication}} and {{Order Parameter Estimation}} of
  {{{\emph{L}}}}1{\textsubscript{0}}-ordered {{FeNi Thin Film Grown}} on a
  {{AuNi Buffer Layer}}, J. Magn. Soc. Jpn. 35~(4) (2011) 370--373.
\newblock \href {https://doi.org/10.3379/msjmag.1106R008}
  {\path{doi:10.3379/msjmag.1106R008}}.

\bibitem{kojima_magnetic_2012}
T.~Kojima, M.~Mizuguchi, T.~Koganezawa, K.~Osaka, M.~Kotsugi, K.~Takanashi,
  Magnetic {{Anisotropy}} and {{Chemical Order}} of {{Artificially Synthesized
  L1}}{\textsubscript{0}}-{{Ordered FeNi Films}} on {{Au}}--{{Cu}}--{{Ni Buffer
  Layers}}, Jpn. J. Appl. Phys. 51~(1R) (2012) 010204.
\newblock \href {https://doi.org/10.1143/JJAP.51.010204}
  {\path{doi:10.1143/JJAP.51.010204}}.

\bibitem{kotsugi_origin_2013}
M.~Kotsugi, M.~Mizuguchi, S.~Sekiya, M.~Mizumaki, T.~Kojima, T.~Nakamura,
  H.~Osawa, K.~Kodama, T.~Ohtsuki, T.~Ohkochi, K.~Takanashi, Y.~Watanabe,
  Origin of strong magnetic anisotropy in {{L1}}{\textsubscript{0}}-{{FeNi}}
  probed by angular-dependent magnetic circular dichroism, J. Magn. Magn.
  Mater. 326 (2013) 235--239.
\newblock \href {https://doi.org/10.1016/j.jmmm.2012.09.008}
  {\path{doi:10.1016/j.jmmm.2012.09.008}}.

\bibitem{ogiwara_magnetization_2013}
M.~Ogiwara, S.~Iihama, T.~Seki, T.~Kojima, S.~Mizukami, M.~Mizuguchi,
  K.~Takanashi, Magnetization damping of an
  {{{\emph{L}}}}1{\textsubscript{0}}-{{FeNi}} thin film with perpendicular
  magnetic anisotropy, Appl. Phys. Lett. 103~(24) (2013) 242409.
\newblock \href {https://doi.org/10.1063/1.4845035}
  {\path{doi:10.1063/1.4845035}}.

\bibitem{ohtsuki_nanoscale_2013}
T.~Ohtsuki, M.~Kotsugi, T.~Ohkochi, S.~Lee, Z.~Horita, K.~Takanashi, Nanoscale
  characterization of {{FeNi}} alloys processed by high-pressure torsion using
  photoelectron emission microscope, J. Appl. Phys. 114~(14) (2013) 143905.
\newblock \href {https://doi.org/10.1063/1.4824372}
  {\path{doi:10.1063/1.4824372}}.

\bibitem{kojima_addition_2014}
T.~Kojima, M.~Mizuguchi, T.~Koganezawa, M.~Ogiwara, M.~Kotsugi, T.~Ohtsuki,
  T.-Y. Tashiro, K.~Takanashi, Addition of {{Co}} to
  {{L1}}{\textsubscript{0}}-ordered {{FeNi}} films: Influences on magnetic
  properties and ordered structures, J. Phys. Appl. Phys. 47~(42) (2014)
  425001.
\newblock \href {https://doi.org/10.1088/0022-3727/47/42/425001}
  {\path{doi:10.1088/0022-3727/47/42/425001}}.

\bibitem{mibu_local_2015}
K.~Mibu, T.~Kojima, M.~Mizuguchi, K.~Takanashi, Local structure and magnetism
  of {{L1}}{\textsubscript{0}}-type {{FeNi}} alloy films with perpendicular
  magnetic anisotropy studied through {\textsuperscript{57}}{{Fe}} nuclear
  probes, J. Phys. Appl. Phys. 48~(20) (2015) 205002.
\newblock \href {https://doi.org/10.1088/0022-3727/48/20/205002}
  {\path{doi:10.1088/0022-3727/48/20/205002}}.

\bibitem{ueno_structural_2015}
T.~Ueno, K.~Saito, N.~Inami, T.~Kojima, M.~Mizuguchi, N.~Miyata, K.~Akutsu,
  M.~Takeda, K.~Takanashi, K.~Ono, Structural and {{Magnetic Depth Profile
  Analysis}} of {{L1}}{\textsubscript{0}} {{FeNi Film}} by {{Polarized Neutron
  Reflectometry}}, in: Proc. 2nd {{Int}}. {{Symp}}. {{Sci}}. {{J-PARC}},
  Journal of the Physical Society of Japan, Tsukuba, Ibaraki, Japan, 2015.
\newblock \href {https://doi.org/10.7566/JPSCP.8.034008}
  {\path{doi:10.7566/JPSCP.8.034008}}.

\bibitem{kojima_growth_2016}
T.~Kojima, M.~Mizuguchi, K.~Takanashi, Growth of
  {{L1}}{\textsubscript{0}}--{{FeNi}} thin films on {{Cu}}(001) single crystal
  substrates using oxygen and gold surfactants, Thin Solid Films 603 (2016)
  348--352.
\newblock \href {https://doi.org/10.1016/j.tsf.2016.02.040}
  {\path{doi:10.1016/j.tsf.2016.02.040}}.

\bibitem{tashiro_fabrication_2018}
T.~Tashiro, M.~Mizuguchi, T.~Kojima, T.~Koganezawa, M.~Kotsugi, T.~Ohtsuki,
  K.~Sato, T.~Konno, K.~Takanashi, Fabrication of
  {{L1}}{\textsubscript{0}}-{{FeNi}} phase by sputtering with rapid thermal
  annealing, J. Alloys Compd. 750 (2018) 164--170.
\newblock \href {https://doi.org/10.1016/j.jallcom.2018.02.318}
  {\path{doi:10.1016/j.jallcom.2018.02.318}}.

\bibitem{ito_epitaxial_2020}
K.~Ito, M.~Hayashida, H.~Masuda, T.~Nishio, S.~Goto, H.~Kura, T.~Koganezawa,
  M.~Mizuguchi, Y.~Shimada, T.~J. Konno, H.~Yanagihara, K.~Takanashi, Epitaxial
  {{L1}}{\textsubscript{0}}-{{FeNi}} films with high degree of order and large
  uniaxial magnetic anisotropy fabricated by denitriding {{FeNiN}} films, Appl.
  Phys. Lett. 116~(24) (2020) 242404.
\newblock \href {https://doi.org/10.1063/5.0011875}
  {\path{doi:10.1063/5.0011875}}.

\bibitem{nishio_fabrication_2021}
T.~Nishio, H.~Kura, K.~Ito, K.~Takanashi, H.~Yanagihara, Fabrication of
  {{L1}}{\textsubscript{0}}-{{FeNi}} films with island structures by nitrogen
  insertion and topotactic extraction for improved coercivity, APL Mater. 9~(9)
  (2021) 091108.
\newblock \href {https://doi.org/10.1063/5.0062692}
  {\path{doi:10.1063/5.0062692}}.

\bibitem{ito_fabrication_2023}
K.~Ito, T.~Ichimura, M.~Hayashida, T.~Nishio, S.~Goto, H.~Kura, R.~Sasaki,
  M.~Tsujikawa, M.~Shirai, T.~Koganezawa, M.~Mizuguchi, Y.~Shimada, T.~J.
  Konno, H.~Yanagihara, K.~Takanashi, Fabrication of
  {{L1}}{\textsubscript{0}}-ordered {{FeNi}} films by denitriding
  {{FeNiN}}(001) and {{FeNiN}}(110) films, J. Alloys Compd. 946 (2023) 169450.
\newblock \href {https://doi.org/10.1016/j.jallcom.2023.169450}
  {\path{doi:10.1016/j.jallcom.2023.169450}}.

\bibitem{nishio_uniaxial_2024}
T.~Nishio, K.~Ito, H.~Kura, K.~Takanashi, H.~Yanagihara, Uniaxial magnetic
  anisotropy of {{L1}}{\textsubscript{0}}-{{FeNi}} films with island structures
  on {{LaAlO}}{\textsubscript{3}}(110) substrates by nitrogen insertion and
  topotactic extraction, J. Alloys Compd. 976 (2024) 172992.
\newblock \href {https://doi.org/10.1016/j.jallcom.2023.172992}
  {\path{doi:10.1016/j.jallcom.2023.172992}}.

\bibitem{takanashi_fabrication_2017}
K.~Takanashi, M.~Mizuguchi, T.~Kojima, T.~Tashiro, Fabrication and
  characterization of {{{\emph{L}}}}1{\textsubscript{0}}-ordered {{FeNi}} thin
  films, J. Phys. Appl. Phys. 50~(48) (2017) 483002.
\newblock \href {https://doi.org/10.1088/1361-6463/aa8ff6}
  {\path{doi:10.1088/1361-6463/aa8ff6}}.

\bibitem{chen_laser-induced_2015}
Z.~Chen, S.~Li, T.~Lai, Laser-induced transient strengthening of coupling in
  {{{\emph{L}}}}1{\textsubscript{0}}-{{FePt}}/{{FeNi}} exchange-spring film, J.
  Phys. Appl. Phys. 48~(14) (2015) 145002.
\newblock \href {https://doi.org/10.1088/0022-3727/48/14/145002}
  {\path{doi:10.1088/0022-3727/48/14/145002}}.

\bibitem{svalov_study_2015}
A.~Svalov, B.~Gonz{\'a}lez~Asensio, A.~Chlenova, P.~Savin, A.~Larra{\~n}aga,
  J.~Gonzalez, G.~Kurlyandskaya, Study of the effect of the deposition rate and
  seed layers on structure and magnetic properties of magnetron sputtered
  {{FeNi}} films, Vacuum 119 (2015) 245--249.
\newblock \href {https://doi.org/10.1016/j.vacuum.2015.05.037}
  {\path{doi:10.1016/j.vacuum.2015.05.037}}.

\bibitem{frisk_resonant_2016}
A.~Frisk, B.~Lindgren, S.~D. Pappas, E.~Johansson, G.~Andersson, Resonant x-ray
  diffraction revealing chemical disorder in sputtered
  {{L1}}{\textsubscript{0}} {{FeNi}} on {{Si}}(0 0 1), J. Phys.: Condens.
  Matter 28~(40) (2016) 406002.
\newblock \href {https://doi.org/10.1088/0953-8984/28/40/406002}
  {\path{doi:10.1088/0953-8984/28/40/406002}}.

\bibitem{frisk_strain_2017}
A.~Frisk, T.~P.~A. Hase, P.~Svedlindh, E.~Johansson, G.~Andersson, Strain
  engineering for controlled growth of thin-film {{FeNi
  L1}}{\textsubscript{0}}, J. Phys. Appl. Phys. 50~(8) (2017) 085009.
\newblock \href {https://doi.org/10.1088/1361-6463/aa5629}
  {\path{doi:10.1088/1361-6463/aa5629}}.

\bibitem{giannopoulos_l10-feni_2018}
G.~Giannopoulos, G.~Barucca, A.~Kaidatzis, V.~Psycharis, R.~Salikhov, M.~Farle,
  E.~Koutsouflakis, D.~Niarchos, A.~Mehta, M.~Scuderi, G.~Nicotra, C.~Spinella,
  S.~Laureti, G.~Varvaro, L1{\textsubscript{0}}-{{FeNi}} films on {{Au-Cu-Ni}}
  buffer-layer: A high-throughput combinatorial study, Sci. Rep. 8~(1) (2018)
  15919.
\newblock \href {https://doi.org/10.1038/s41598-018-34296-9}
  {\path{doi:10.1038/s41598-018-34296-9}}.

\bibitem{nguyen_ordered_2023}
V.~Q. Nguyen, B.-H. Jun, Y.-B. Chun, J.~H. Lee, Ordered
  {{L1}}{\textsubscript{0}}-{{FeNi}} (111) epitaxial thin film on
  {{Al}}{\textsubscript{2}}{{O}}{\textsubscript{3}} (0001) substrate:
  {{Molecular}} beam epitaxy growth and characterizations, Thin Solid Films 780
  (2023) 139962.
\newblock \href {https://doi.org/10.1016/j.tsf.2023.139962}
  {\path{doi:10.1016/j.tsf.2023.139962}}.

\bibitem{mandal_l10_2023}
S.~Mandal, M.~Debata, P.~Sengupta, S.~Basu, L1{\textsubscript{0}} {{FeNi}}: A
  promising material for next generation permanent magnets, Crit. Rev. Solid
  State Mater. Sci. 48~(6) (2023) 703--725.
\newblock \href {https://doi.org/10.1080/10408436.2022.2107484}
  {\path{doi:10.1080/10408436.2022.2107484}}.

\bibitem{sakamaki_effect_2013}
M.~Sakamaki, K.~Amemiya, Effect of structural strain on magnetic anisotropy
  energy of each element in alternately layered {{FeNi}} thin films, Phys. Rev.
  B 87~(1) (2013) 014428.
\newblock \href {https://doi.org/10.1103/PhysRevB.87.014428}
  {\path{doi:10.1103/PhysRevB.87.014428}}.

\bibitem{kojima_feni_2014}
T.~Kojima, M.~Ogiwara, M.~Mizuguchi, M.~Kotsugi, T.~Koganezawa, T.~Ohtsuki,
  T.-Y. Tashiro, K.~Takanashi, Fe--{{Ni}} composition dependence of magnetic
  anisotropy in artificially fabricated {{L1}}{\textsubscript{0}}-ordered
  {{FeNi}} films, J. Phys. Condens. Matter 26~(6) (2014) 064207.
\newblock \href {https://doi.org/10.1088/0953-8984/26/6/064207}
  {\path{doi:10.1088/0953-8984/26/6/064207}}.

\bibitem{lewis_magnete_2014}
L.~H. Lewis, F.~E. Pinkerton, N.~Bordeaux, A.~Mubarok, E.~Poirier, J.~I.
  Goldstein, R.~Skomski, K.~Barmak, De {{Magnete}} et {{Meteorite}}:
  {{Cosmically Motivated Materials}}, IEEE Magn. Lett. 5 (2014) 1--4.
\newblock \href {https://doi.org/10.1109/LMAG.2014.2312178}
  {\path{doi:10.1109/LMAG.2014.2312178}}.

\bibitem{yang_revision_1996}
C.~W. Yang, D.~B. Williams, J.~I. Goldstein, A revision of the {{Fe-Ni}} phase
  diagram at low temperatures ({$<$}400 {$^\circ$}{{C}}), J. Phase Equilibria
  17~(6) (1996) 522--531.
\newblock \href {https://doi.org/10.1007/BF02665999}
  {\path{doi:10.1007/BF02665999}}.

\bibitem{woodgate_revisiting_2023}
C.~D. Woodgate, C.~E. Patrick, L.~H. Lewis, J.~B. Staunton, Revisiting
  {{N{\'e}el}} 60 years on: {{The}} magnetic anisotropy of
  {{L1}}{\textsubscript{0}} {{FeNi}} (tetrataenite), J. Appl. Phys. 134~(16)
  (2023) 163905.
\newblock \href {https://doi.org/10.1063/5.0169752}
  {\path{doi:10.1063/5.0169752}}.

\bibitem{wu_spinorbit_1999}
R.~Wu, A.~J. Freeman, Spin--orbit induced magnetic phenomena in bulk metals and
  their surfaces and interfaces, J. Magn. Magn. Mater. 200~(1--3) (1999)
  498--514.
\newblock \href {https://doi.org/10.1016/S0304-8853(99)00351-0}
  {\path{doi:10.1016/S0304-8853(99)00351-0}}.

\bibitem{ravindran_large_2001}
P.~Ravindran, A.~Kjekshus, H.~Fjellv{\aa}g, P.~James, L.~Nordstr{\"o}m,
  B.~Johansson, O.~Eriksson, Large magnetocrystalline anisotropy in bilayer
  transition metal phases from first-principles full-potential calculations,
  Phys. Rev. B 63~(14) (2001) 144409.
\newblock \href {https://doi.org/10.1103/PhysRevB.63.144409}
  {\path{doi:10.1103/PhysRevB.63.144409}}.

\bibitem{miura_origin_2013}
Y.~Miura, S.~Ozaki, Y.~Kuwahara, M.~Tsujikawa, K.~Abe, M.~Shirai, The origin of
  perpendicular magneto-crystalline anisotropy in {{L1}}{\textsubscript{0}}
  {{FeNi}} under tetragonal distortion, J. Phys. Condens. Matter 25~(10) (2013)
  106005.
\newblock \href {https://doi.org/10.1088/0953-8984/25/10/106005}
  {\path{doi:10.1088/0953-8984/25/10/106005}}.

\bibitem{edstrom_electronic_2014}
A.~Edstr{\"o}m, J.~Chico, A.~Jakobsson, A.~Bergman, J.~Rusz, Electronic
  structure and magnetic properties of {{L1}}{\textsubscript{0}} binary alloys,
  Phys. Rev. B 90~(1) (2014) 014402.
\newblock \href {https://doi.org/10.1103/PhysRevB.90.014402}
  {\path{doi:10.1103/PhysRevB.90.014402}}.

\bibitem{manchanda_transition-metal_2014}
P.~Manchanda, R.~Skomski, N.~Bordeaux, L.~H. Lewis, A.~Kashyap,
  Transition-metal and metalloid substitutions in
  {{L1}}{\textsubscript{0}}-ordered {{FeNi}}, J. Appl. Phys. 115~(17) (2014)
  17A710.
\newblock \href {https://doi.org/10.1063/1.4862722}
  {\path{doi:10.1063/1.4862722}}.

\bibitem{tian_density_2019}
L.-Y. Tian, H.~Lev{\"a}m{\"a}ki, O.~Eriksson, K.~Kokko, {\'A}.~Nagy, E.~K.
  {D{\'e}lczeg-Czirj{\'a}k}, L.~Vitos, Density {{Functional Theory}}
  description of the order-disorder transformation in {{Fe-Ni}}, Sci. Rep.
  9~(1) (2019) 8172.
\newblock \href {https://doi.org/10.1038/s41598-019-44506-7}
  {\path{doi:10.1038/s41598-019-44506-7}}.

\bibitem{tian_pressure_2020}
L.-Y. Tian, O.~Eriksson, L.~Vitos, Pressure effect on the order--disorder
  transformation in {{L1}}{\textsubscript{0}} {{FeNi}}, Sci. Rep. 10~(1) (2020)
  14766.
\newblock \href {https://doi.org/10.1038/s41598-020-71551-4}
  {\path{doi:10.1038/s41598-020-71551-4}}.

\bibitem{tian_alloying_2021}
L.-Y. Tian, O.~Gutfleisch, O.~Eriksson, L.~Vitos, Alloying effect on the
  order--disorder transformation in tetragonal {{FeNi}}, Sci. Rep. 11~(1)
  (2021) 5253.
\newblock \href {https://doi.org/10.1038/s41598-021-84482-5}
  {\path{doi:10.1038/s41598-021-84482-5}}.

\bibitem{tuvshin_fenin_2021}
D.~Tuvshin, T.~Tsevelmaa, S.~Hong, D.~Odkhuu, {{Fe}}-{{Ni}}-{{N}} based alloys
  as rare-earth free high-performance permanent magnet across {$\alpha$}'' to
  {{L1}}{\textsubscript{0}} phase transition: {{A}} theoretical insight, Acta
  Mater. 210 (2021) 116807.
\newblock \href {https://doi.org/10.1016/j.actamat.2021.116807}
  {\path{doi:10.1016/j.actamat.2021.116807}}.

\bibitem{si_effect_2022}
M.~Si, A.~Izardar, C.~Ederer, Effect of chemical disorder on the magnetic
  anisotropy in {{L1}}{\textsubscript{0}} {{FeNi}} from first-principles
  calculations, Phys. Rev. Res. 4~(3) (2022) 033161.
\newblock \href {https://doi.org/10.1103/PhysRevResearch.4.033161}
  {\path{doi:10.1103/PhysRevResearch.4.033161}}.

\bibitem{qiao_effect_2023}
Z.~Qiao, M.~Tsujikawa, M.~Shirai, The effect of chemical disorder on magnetic
  properties of {{FeNi}} and
  {{Fe}}{\textsubscript{2}}{{Ni}}{\textsubscript{2}}{{N}} alloys, J. Magn.
  Magn. Mater. 568 (2023) 170362.
\newblock \href {https://doi.org/10.1016/j.jmmm.2023.170362}
  {\path{doi:10.1016/j.jmmm.2023.170362}}.

\bibitem{yamashita_finite-temperature_2023}
S.~Yamashita, A.~Sakuma, Finite-temperature second-order perturbation analysis
  of magnetocrystalline anisotropy energy of {{L1}}{\textsubscript{0}}-type
  ordered alloys, Phys. Rev. B 108~(5) (2023) 054411.
\newblock \href {https://doi.org/10.1103/PhysRevB.108.054411}
  {\path{doi:10.1103/PhysRevB.108.054411}}.

\bibitem{simonetti_study_2010}
S.~Simonetti, G.~Brizuela, A.~Juan, Study of the adsorption, electronic
  structure and bonding of {{C}}{\textsubscript{2}}{{H}}{\textsubscript{4 }}on
  the {{FeNi}}(111) surface, Appl. Surf. Sci. 256~(21) (2010) 6459--6465.
\newblock \href {https://doi.org/10.1016/j.apsusc.2010.04.035}
  {\path{doi:10.1016/j.apsusc.2010.04.035}}.

\bibitem{he_first-principles_2015}
Y.-B. He, J.-F. Jia, H.-S. Wu, First-{{Principles Investigation}} of the
  {{Molecular Adsorption}} and {{Dissociation}} of {{Hydrazine}} on
  {{Ni}}--{{Fe Alloy Surfaces}}, J. Phys. Chem. C 119~(16) (2015) 8763--8774.
\newblock \href {https://doi.org/10.1021/acs.jpcc.5b01605}
  {\path{doi:10.1021/acs.jpcc.5b01605}}.

\bibitem{marciniak_dft_2022}
J.~Marciniak, W.~Marciniak, M.~Werwi{\'n}ski, {{DFT}} calculation of intrinsic
  properties of magnetically hard phase {{L1}}{\textsubscript{0}} {{FePt}}, J.
  Magn. Magn. Mater. 556 (2022) 169347.
\newblock \href {https://doi.org/10.1016/j.jmmm.2022.169347}
  {\path{doi:10.1016/j.jmmm.2022.169347}}.

\bibitem{marciniak_l10_2023}
J.~Marciniak, M.~Werwi{\'n}ski, L1{\textsubscript{0}} {{FePt}} thin films with
  tilted and in-plane magnetic anisotropy: {{A}} first-principles study, Phys.
  Rev. B 108~(21) (2023) 214406.
\newblock \href {https://doi.org/10.1103/PhysRevB.108.214406}
  {\path{doi:10.1103/PhysRevB.108.214406}}.

\bibitem{albrecht_magnetic_2005}
M.~Albrecht, G.~Hu, I.~L. Guhr, T.~C. Ulbrich, J.~Boneberg, P.~Leiderer,
  G.~Schatz, Magnetic multilayers on nanospheres, Nat. Mater. 4~(3) (2005)
  203--206.
\newblock \href {https://doi.org/10.1038/nmat1324}
  {\path{doi:10.1038/nmat1324}}.

\bibitem{wang_tilting_2005}
J.-P. Wang, Tilting for the top, Nat. Mater. 4~(3) (2005) 191--192.
\newblock \href {https://doi.org/10.1038/nmat1344}
  {\path{doi:10.1038/nmat1344}}.

\bibitem{zha_pseudo_2009}
C.~L. Zha, J.~Persson, S.~Bonetti, Y.~Y. Fang, J.~{\AA}kerman, Pseudo spin
  valves based on {{L1}}{\textsubscript{0}} (111)-oriented {{FePt}} fixed
  layers with tilted anisotropy, Appl. Phys. Lett. 94~(16) (2009) 163108.
\newblock \href {https://doi.org/10.1063/1.3123003}
  {\path{doi:10.1063/1.3123003}}.

\bibitem{hsu_situ_2000}
Y.-N. Hsu, S.~Jeong, D.~Lambeth, D.~Laughlin, In situ ordering of {{FePt}} thin
  films by using {{Ag}}/{{Si}} and
  {{Ag}}/{{Mn}}{\textsubscript{3}}{{Si}}/{{Ag}}/{{Si}} templates, IEEE Trans.
  Magn. 36~(5) (2000) 2945--2947.
\newblock \href {https://doi.org/10.1109/20.908636}
  {\path{doi:10.1109/20.908636}}.

\bibitem{shima_preparation_2002}
T.~Shima, K.~Takanashi, Y.~K. Takahashi, K.~Hono, Preparation and magnetic
  properties of highly coercive {{FePt}} films, Appl. Phys. Lett. 81~(6) (2002)
  1050--1052.
\newblock \href {https://doi.org/10.1063/1.1498504}
  {\path{doi:10.1063/1.1498504}}.

\bibitem{ohtake_l10_2012}
M.~Ohtake, S.~Ouchi, F.~Kirino, M.~Futamoto, L10 ordered phase formation in
  {{FePt}}, {{FePd}}, {{CoPt}}, and {{CoPd}} alloy thin films epitaxially grown
  on {{MgO}}(001) single-crystal substrates, J. Appl. Phys. 111~(7) (2012)
  07A708.
\newblock \href {https://doi.org/10.1063/1.3672856}
  {\path{doi:10.1063/1.3672856}}.

\bibitem{sepehri-amin_microstructure_2017}
H.~{Sepehri-Amin}, H.~Iwama, T.~Ohkubo, T.~Shima, K.~Hono, Microstructure and
  in-plane component of {{L1}}{\textsubscript{0}}-{{FePt}} films deposited on
  {{MgO}} and {{MgAl}}{\textsubscript{2}}{{O}}{\textsubscript{4}} substrates,
  Scr. Mater. 130 (2017) 247--251.
\newblock \href {https://doi.org/10.1016/j.scriptamat.2016.12.018}
  {\path{doi:10.1016/j.scriptamat.2016.12.018}}.

\bibitem{wu_atomic-scale_2022}
K.~Wu, X.~Fu, W.~Zhu, X.~Huang, Atomic-scale investigation on the origin of
  in-plane variants in {{L1}}{\textsubscript{0}}-{{FePt}} nanoparticles
  embedded in a single-crystalline {{MgO}} matrix, J. Appl. Phys. 132~(17)
  (2022) 175305.
\newblock \href {https://doi.org/10.1063/5.0109411}
  {\path{doi:10.1063/5.0109411}}.

\bibitem{koepernik_full-potential_1999}
K.~Koepernik, H.~Eschrig, Full-potential nonorthogonal local-orbital
  minimum-basis band-structure scheme, Phys. Rev. B 59~(3) (1999) 1743--1757.
\newblock \href {https://doi.org/10.1103/PhysRevB.59.1743}
  {\path{doi:10.1103/PhysRevB.59.1743}}.

\bibitem{opahle_full-potential_1999}
I.~Opahle, K.~Koepernik, H.~Eschrig, Full-potential band-structure calculation
  of iron pyrite, Phys. Rev. B 60~(20) (1999) 14035.
\newblock \href {https://doi.org/10.1103/PhysRevB.60.14035}
  {\path{doi:10.1103/PhysRevB.60.14035}}.

\bibitem{perdew_generalized_1996}
J.~P. Perdew, K.~Burke, M.~Ernzerhof, Generalized gradient approximation made
  simple, Phys. Rev. Lett. 77~(18) (1996) 3865--3868.
\newblock \href {https://doi.org/10.1103/PhysRevLett.77.3865}
  {\path{doi:10.1103/PhysRevLett.77.3865}}.

\bibitem{marciniak_structural_2023}
W.~Marciniak, M.~Werwi{\'n}ski, Structural and magnetic properties of
  {{Fe-Co-C}} alloys with tetragonal deformation: {{A}} first-principles study,
  Phys. Rev. B 108~(21) (2023) 214433.
\newblock \href {https://doi.org/10.1103/PhysRevB.108.214433}
  {\path{doi:10.1103/PhysRevB.108.214433}}.

\bibitem{perdew_accurate_1992}
J.~P. Perdew, Y.~Wang, Accurate and simple analytic representation of the
  electron-gas correlation energy, Phys. Rev. B 45~(23) (1992) 13244--13249.
\newblock \href {https://doi.org/10.1103/PhysRevB.45.13244}
  {\path{doi:10.1103/PhysRevB.45.13244}}.

\bibitem{liechtenstein_local_1987}
A.~I. Liechtenstein, M.~I. Katsnelson, V.~P. Antropov, V.~A. Gubanov, Local
  spin density functional approach to the theory of exchange interactions in
  ferromagnetic metals and alloys, J. Magn. Magn. Mater. 67~(1) (1987) 65--74.
\newblock \href {https://doi.org/10.1016/0304-8853(87)90721-9}
  {\path{doi:10.1016/0304-8853(87)90721-9}}.

\bibitem{wang_validity_1996}
X.~Wang, D.-s. Wang, R.~Wu, A.~J. Freeman, Validity of the force theorem for
  magnetocrystalline anisotropy, J. Magn. Magn. Mater. 159~(3) (1996) 337--341.
\newblock \href {https://doi.org/10.1016/0304-8853(95)00936-1}
  {\path{doi:10.1016/0304-8853(95)00936-1}}.

\bibitem{mulliken_electronic_1955}
R.~S. Mulliken, Electronic {{Population Analysis}} on {{LCAO}}--{{MO Molecular
  Wave Functions}}. {{I}}, J. Chem. Phys. 23~(10) (1955) 1833--1840.
\newblock \href {https://doi.org/10.1063/1.1740588}
  {\path{doi:10.1063/1.1740588}}.

\bibitem{momma_vesta_2008}
K.~Momma, F.~Izumi, {{{\emph{VESTA}}}} : A three-dimensional visualization
  system for electronic and structural analysis, J. Appl. Crystallogr. 41~(3)
  (2008) 653--658.
\newblock \href {https://doi.org/10.1107/S0021889808012016}
  {\path{doi:10.1107/S0021889808012016}}.

\bibitem{dannenberg_surface_2009}
A.~Dannenberg, M.~E. Gruner, A.~Hucht, P.~Entel, Surface energies of
  stoichiometric {{FePt}} and {{CoPt}} alloys and their implications for
  nanoparticle morphologies, Phys. Rev. B 80~(24) (2009) 245438.
\newblock \href {https://doi.org/10.1103/PhysRevB.80.245438}
  {\path{doi:10.1103/PhysRevB.80.245438}}.

\bibitem{szunyogh_oscillatory_1997}
L.~Szunyogh, B.~{\'U}jfalussy, C.~Blaas, U.~Pustogowa, C.~Sommers,
  P.~Weinberger, Oscillatory behavior of the magnetic anisotropy energy in
  {{Cu}}(100)/{{Co}}{\textsubscript{n}} multilayer systems, Phys. Rev. B
  56~(21) (1997) 14036--14044.
\newblock \href {https://doi.org/10.1103/PhysRevB.56.14036}
  {\path{doi:10.1103/PhysRevB.56.14036}}.

\bibitem{zhang_electric-field_2009}
H.~Zhang, M.~Richter, K.~Koepernik, I.~Opahle, F.~Tasn{\'a}di, H.~Eschrig,
  Electric-field control of surface magnetic anisotropy: A density functional
  approach, New J. Phys. 11~(4) (2009) 043007.
\newblock \href {https://doi.org/10.1088/1367-2630/11/4/043007}
  {\path{doi:10.1088/1367-2630/11/4/043007}}.

\bibitem{blanco-rey_large_2021}
M.~{Blanco-Rey}, P.~Perna, A.~Gudin, J.~M. Diez, A.~Anad{\'o}n,
  P.~{Olleros-Rodr{\'i}guez}, L.~De~Melo~Costa, M.~Valvidares, P.~Gargiani,
  A.~{Guedeja-Marron}, M.~Cabero, M.~Varela, C.~{Garc{\'i}a-Fern{\'a}ndez},
  M.~M. Otrokov, J.~Camarero, R.~Miranda, A.~Arnau, J.~I. Cerd{\'a}, Large
  {{Perpendicular Magnetic Anisotropy}} in {{Nanometer-Thick Epitaxial
  Graphene}}/{{Co}}/{{Heavy Metal Heterostructures}} for
  {{Spin}}--{{Orbitronics Devices}}, ACS Appl. Nano Mater. 4~(5) (2021)
  4398--4408.
\newblock \href {https://doi.org/10.1021/acsanm.0c03364}
  {\path{doi:10.1021/acsanm.0c03364}}.

\bibitem{chang_voltage-controlled_2021}
P.-H. Chang, W.~Fang, T.~Ozaki, K.~D. Belashchenko, Voltage-controlled magnetic
  anisotropy in antiferromagnetic {{MgO-capped MnPt}} films, Phys. Rev. Mater.
  5~(5) (2021) 054406.
\newblock \href {https://doi.org/10.1103/PhysRevMaterials.5.054406}
  {\path{doi:10.1103/PhysRevMaterials.5.054406}}.

\bibitem{cinal_magnetic_2022}
M.~Cinal, Magnetic anisotropy and orbital magnetic moment in {{Co}} films and
  {{Co}}/{{X}} bilayers ({{X}} = {{Pd}} and {{Pt}}), Phys. Rev. B 105~(10)
  (2022) 104403.
\newblock \href {https://doi.org/10.1103/PhysRevB.105.104403}
  {\path{doi:10.1103/PhysRevB.105.104403}}.

\bibitem{przybylski_oscillatory_2012}
M.~Przybylski, M.~D{\k a}browski, U.~Bauer, M.~Cinal, J.~Kirschner, Oscillatory
  magnetic anisotropy due to quantum well states in thin ferromagnetic films
  (invited), J. Appl. Phys. 111~(7) (2012) 07C102.
\newblock \href {https://doi.org/10.1063/1.3670498}
  {\path{doi:10.1063/1.3670498}}.

\bibitem{bruno_tight-binding_1989}
P.~Bruno, Tight-binding approach to the orbital magnetic moment and
  magnetocrystalline anisotropy of transition-metal monolayers, Phys. Rev. B
  39~(1) (1989) 865.
\newblock \href {https://doi.org/10.1103/PhysRevB.39.865}
  {\path{doi:10.1103/PhysRevB.39.865}}.

\bibitem{ke_intersublattice_2019}
L.~Ke, Intersublattice magnetocrystalline anisotropy using a realistic
  tight-binding method based on maximally localized {{Wannier}} functions,
  Phys. Rev. B 99~(5) (2019) 054418.
\newblock \href {https://doi.org/10.1103/PhysRevB.99.054418}
  {\path{doi:10.1103/PhysRevB.99.054418}}.

\bibitem{kovacs_micromagnetic_2017}
A.~Kovacs, H.~Ozelt, J.~Fischbacher, T.~Schrefl, A.~Kaidatzis, R.~Salikhov,
  M.~Farle, G.~Giannopoulos, D.~Niarchos, Micromagnetic {{Simulations}} for
  {{Coercivity Improvement}} through {{Nano-Structuring}} of {{Rare-Earth Free
  L1}}{\textsubscript{0}}-{{FeNi Magnets}}, IEEE Trans. Magn. 53~(11) (2017)
  1--5.
\newblock \href {https://doi.org/10.1109/TMAG.2017.2701418}
  {\path{doi:10.1109/TMAG.2017.2701418}}.

\bibitem{skomski_magnetic_2016}
R.~Skomski, J.~Coey, Magnetic anisotropy --- {{How}} much is enough for a
  permanent magnet?, Scr. Mater. 112 (2016) 3--8.
\newblock \href {https://doi.org/10.1016/j.scriptamat.2015.09.021}
  {\path{doi:10.1016/j.scriptamat.2015.09.021}}.

\end{thebibliography}

\end{document}